# A boundary integral based particle initialization algorithm for Smooth Particle Hydrodynamics

Parikshit Boregowda*, Gui-Rong Liu

*College of Engineering and Applied Science, University of Cincinnati, Cincinnati, Ohio, USA*

## Abstract

Algorithms for initializing particle distribution in SPH simulations are important for improving simulation accuracy. However, no such algorithms exist for boundary integral SPH models, which can model complex geometries without requiring layers of virtual particles. This study introduces the Boundary Integral based Particle Initialization (BIPI) algorithm. It employs a particle packing algorithm meticulously designed to redistribute particles to fit the geometry boundary. The BIPI algorithm directly utilizes the geometry's boundary information using the SPH boundary integral formulation. Special consideration is given to particles adjacent to the boundary to prevent artificial volume compression. The BIPI algorithm can hence generate a particle distribution with reduced concentration gradients for domains with complex geometrical shapes. Finally, several examples are presented to demonstrate the effectiveness of the proposed algorithm, including the application of the BIPI algorithm in flow problems.

## 1 Introduction

Smooth Particle Hydrodynamics (SPH) is a mesh-free numerical method that relies on a kernel function, also called a smoothing kernel, $W$, to approximate field variables[1]. The construction of an appropriate smoothing kernel theoretically allows the determination of higher-order derivatives of a field variable[2]. Additionally, the construction of the smoothing kernel allows for kernel-consistent conditions, which include $\int W = 1$ and $\int \partial W = 0$. However, practical computational simulations use summations for finite SPH particles, represented as $\sum W$ and $\sum \partial W$. The discrete form leads to deviations from the kernel consistent conditions, introducing errors in particle approximations[3–5]. While efforts to enhance SPH accuracy using normalization techniques exist, they inadvertently compromise the conservative nature of the SPH formulation[6,7].

Conservative SPH formulations, crucial for momentum and energy conservation, strictly adhere to kernel consistency[8]. The uniform distribution of particles notably influences the accuracy of the kernel consistent SPH formulations. Thus, methodologies in the literature that improve particle distribution, such as particle packing for initialization[9–11], particle shifting for flow evolution[12–14], and ALE SPH formulations[15], directly impact simulation quality.

Another significant challenge in SPH emerges near domain boundaries when a smoothing kernel is truncated. Traditionally, addressing this issue involves adding layers of virtual particles[16,17], a progressively complex approach for intricate boundary shapes. Alternatively, considering the boundary integral terms of SPH formulations has gained attention[18–21], offering a promising avenue to overcome the limitations of virtual particles.

In an effort to achieve a uniform particle distribution before starting a simulation, Monaghan[22] introduced a method involving a high damping term to initiate simulations, allowing particles to settle into equilibrium positions. However, this approach significantly increases computational demands and has been observed to lead to particle resettlement at the onset of the actual numerical simulation. Subsequently, Colagrossi et al. [9] introduced an effective particle packing scheme, emphasizing the importance of reducing particle resettlement under static conditions. Considerations for particle resettlement have also been explored by Litvinov et al.[23], demonstrating enhanced convergence characteristics of Smoothed Particle Hydrodynamics (SPH) formulations for uniform particle distribution. More recently, Negi et al. [10] presented an improved particle packing algorithm tailored for complex shapes by integrating the approaches of [9] and [24]. However, the methodologies of [9,10,22,25] use the virtual particles to model boundaries, making them incompatible with the boundary integral SPH model. In fact, algorithms capable of packing particles into computational geometry without needing layers of virtual particles benefit all SPH research.

Moreover, our attempts to modify existing packing algorithms, incorporating damping terms to achieve stability near the boundary with a purely boundary integral model, have proven challenging. While the algorithm of Jiang et al. [24] can pack particles without modeling ghost particles, their algorithm is designed specifically for blue noise sampling. Additionally, modifying their algorithm would still lead to particles on the boundary – however, in SPH simulations, we require the particles to be strictly inside the boundary. A more promising algorithm to initialize particle distribution without dealing with virtual or ghost particle layers for boundary models is the CAD BPG algorithm and its variants [26,27]. However, in addition to requiring level set information, their algorithm uses traditional SPH formulation, which does not address the inaccuracies of the truncated kernel. Thus, a compelling need exists to develop a dedicated particle initialization algorithm for boundary integral SPH models by directly employing the boundary integral formulations.

## 2 Boundary modeling in SPH

### 2.1 The fundamental boundary integral formulation

Boundary integral formulations of Smooth Particle Hydrodynamics allow for handling boundaries by simply accounting for the otherwise neglected boundary integral in SPH derivative formulation, given by -

$$\partial_i f(\pmb{x}) = - \int_{\partial(\Omega \cap \Omega_w)} f(\pmb{x}')\, W(\pmb{x} - \pmb{x}', \kappa h)\, \underline{\hat{n}}_i \, dS' - \int_{\Omega \cap \Omega_w} f(\pmb{x}')\, \partial_i W(\pmb{x} - \pmb{x}', \kappa h)\, dV' \quad (1)$$

In (1), the smoothing domain $\Omega_w$ is truncated by the boundary of the domain $\Omega$, with the field variable denoted by $f$, and its partial derivative in the $i^{th}$ direction denoted by $\partial_i f(\pmb{x})$ using the smoothing kernel $W(\pmb{x} - \pmb{x}', \kappa h)$, the kernel's partial derivative $\partial_i W(\pmb{x} - \pmb{x}', \kappa h)$, and the surface inward normal $\underline{\hat{n}}$. Also, all derivatives in (1) are calculated with regards to the variable $\pmb{x}$. Additionally, $h$ denotes the smoothing length, and $\kappa h$ represents the radius of the smoothing domain $\Omega_w$ for the smoothing kernel used in the analysis. The boundary integral in (1) is zero only for untruncated smoothing kernel and is otherwise non-zero.

As demonstrated in our previous work [4], a consistent analysis of the integral formulation (1) reveals that the necessary smoothing kernel properties are still violated at the boundary due to the truncated kernel (see Figure 1). For kernel consistency, the gradient of the field is modified in its discrete form, as,

$$\pmb{\nabla} f_a = \frac{1}{\gamma_a}\left(\sum_b f_b\, \pmb{\nabla}_a W_{ab}\, \frac{m_b}{\rho_b} - \sum_s \int_{\partial(\Omega \cap \Omega_w)_s} f(\pmb{x}'_s) W_{as'} \underline{\hat{\pmb{n}}}_s\, dS'\right) \quad (2)$$

Where, $\gamma_a = \int_{\Omega \cap \Omega_w} W(\pmb{x} - \pmb{x}', h)\, dV'$ and $\underline{\hat{\pmb{n}}}_s$ is the inward pointing surface normal for the boundary element $s$. Unlike in (1), in (2) we represent the kernel derivative with respect to the particle, giving us $\pmb{\nabla}_a W_{ab} = -\pmb{\nabla} W_{ab}$.

The boundary integral in (2) can be calculated in various ways. Here, we approximate the boundary integral as:

$$\int_{\partial(\Omega \cap \Omega_w)_s} f(\mathbf{x}'_s) W_{as'} \underline{\hat{\mathbf{n}}}_s \, dS' = f_s \int_{\partial(\Omega \cap \Omega_w)_s} W_{as'} \underline{\hat{\mathbf{n}}}_s \, dS' = f_s \boldsymbol{\nabla} \gamma_{as} \qquad (3)$$

In (3), $f(\mathbf{x}'_s)$ is approximated to be a constant $f_s$ over the boundary element in consideration and moved outside the boundary integral. One-dimensional boundary elements are represented in Figure 1b, with black circles representing the nodes of the boundary element and a black star indicating its centroid. If there are considerable variations in $f(\mathbf{x}'_s)$ within a boundary element, the element would need to be split further into additional segments in (2) before approximating the boundary integral with (3). In this study, we automatically adjust the boundary length ($dx_s$) based on the ratio of particle spacing $dx_r$ by ensuring that $\frac{dx_r}{dx_s} > 1$.

Finally, $\gamma_a$ and $\boldsymbol{\nabla}\gamma_{as}$ is determined analytically for a smoothing kernel. The analytical calculations discussed in Leroy et al. [28] and Feldman et al. [29] for the 2D Quintic Wendland kernel are adopted in the work here. Additionally, to effectively use boundary integral formulations for modeling complex shapes, we employ the no boundary particles (NBP) approach, where the domain boundary is directly represented by line segments (in 2D) and triangles (in 3D).

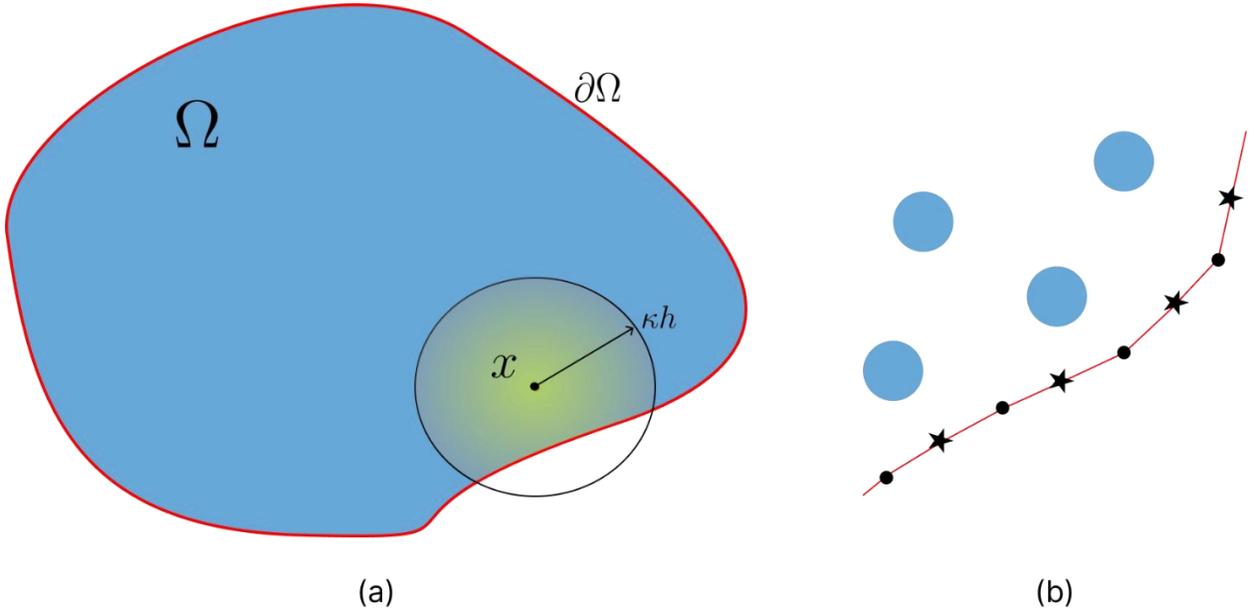

*Figure 1 (a) The smoothing domain $\Omega_w$ at x is truncated by the boundary of the domain $\Omega$, (b) the discretized domain with boundary segments and particles next to the boundary. Here $\Omega$ is the computational domain, $\partial\Omega$ the domain boundary, $\kappa h$ represents the radius of the smoothing domain $\Omega_w$, and $\mathbf{x}$ is the center of kernel.*

## 2.2 Measuring uniform distribution

Achieving a uniform particle distribution is crucial in minimizing the error associated with calculating SPH approximations. For uniform distribution, the terms $\sum_b V_b W_{ab}$ and $\sum_b \nabla_a W_{ab} V_b$ need to be 1 and 0, respectively. This also implies that the uniformity of particle distribution for some complex geometry depends solely on the smoothing kernel and the smoothing length used.

Now, if we employ the boundary integral model of SPH, a constant field, and its gradient is given by,

$$C_a = \frac{\sum_b V_b W_{ab}}{\gamma_a} \qquad (4)$$

and,

$$\nabla C_a = \frac{1}{\gamma_a}\left(\sum_b \nabla_a W_{ab} V_b - \sum_s \nabla \gamma_{as}\right) \qquad (5)$$

Now, to define uniform particle distribution, we would require,

$$C_a \to 1 \qquad (6)$$

$$\nabla C_a \to \mathbf{0} \qquad (7)$$

The magnitude of error in achieving these limiting values, (6) and (7), is higher for truncated kernels compared to untruncated kernels, as discussed in [4]. It is important to note that achieving exact equality for (6) and (7) is not feasible due to its discretized form. However, the primary objective is to reduce the error in approaching these values to improve the accuracy of practical simulations while utilizing the conservative form of SPH.

## 3 A Boundary Integral based Particle Initialization (BIPI) algorithm

### 3.1 Generating particles from computational geometry file: Step 1

The initial step in any particle initialization algorithm involves generating particles from the geometry file. This process can be executed in two ways: by substituting the underlying Cartesian grid (voxel in 3D and pixel in 2D) of the geometry's interior with particles or by meshing the geometry's volume and subsequently replacing each volume element with a particle. The former approach is commonly employed within the SPH community [30,31].

In our approach, we leverage the geometry's boundary information to generate particles relative to an underlying Cartesian grid, and we directly utilize the boundary elements for calculating the boundary integral in equation (3). This data is readily available in the STL file format for 3D geometries with triangles of sufficient size tolerance, while for 2D geometries, line elements are utilized to represent the boundary (as shown in Figure 2). The orientation of the surface normal is important, as it distinguishes the interior from the exterior of the boundary.

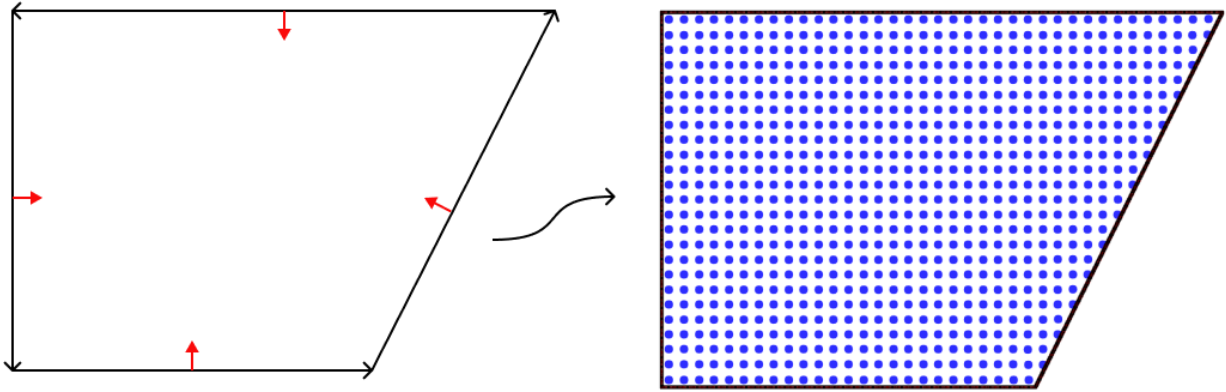

*Figure 2 A point in polygon algorithm is used to generate one particle per cartesian cell using boundary edges and boundary sense to differentiate the interior from the exterior of the polygon.*

### 3.2 Shift particles using the concentration gradient: Step 2

The next step in the BIPI algorithm is to pack particles along the boundary while also aiming to achieve a uniform distribution. To achieve this, we start with the symmetric form of SPH gradient formulation (2),

$$\boldsymbol{\nabla} f_{a_+} = \frac{1}{\gamma_a}\left(\sum_b (f_a + f_b)\,\boldsymbol{\nabla}_a W\, V_b - \sum_s (f_a + f_s)\boldsymbol{\nabla}\gamma_{as}\right) \qquad (8)$$

This form of gradient is used for representing pressure ($p$) in SPH application to fluid flow due to its conservative nature and can be derived by simply adding $\boldsymbol{\nabla} 1$ to $\boldsymbol{\nabla} f_a$ using (2), as $\boldsymbol{\nabla} 1$ in the integral form is 0.

To pack the particles, we start with a simple momentum equation following the works of [9,10], but with only a pressure gradient term, as $\ddot{x}_a = -\boldsymbol{\nabla} p_a/\rho_a$. Assuming all particles have the same density ($\rho_a = \rho_0$) and pressure ($p_a = p_0$), with pressure at the boundary ($p_s$) equal to that inside ($p_0$), we aim for particles to reach equilibrium, which simplifies (8) for pressure gradient $\nabla p_{a_+}$ to,

$$\ddot{x}_a = -\frac{2\,p_0}{\gamma_a\,\rho_0}\left(\sum_b \nabla_a W\, V_b - \sum_s \nabla \gamma_{as}\right) = -\frac{2\,p_0}{\rho_0}\,\nabla C_a \tag{9}$$

A stable configuration is achieved when the acceleration in equation (9) is zero, necessitating $\nabla C_a$ to approach zero. In other words, equation (9) dictates particle movement to minimize the concentration gradient, akin to a Particle Shifting Technique (PST) as follows,

$$\widetilde{\delta x}_a = -D_a\,\nabla C_a \tag{10}$$

Where,

$$D_a = (\delta t)^2\,\frac{2\,p_0}{\rho_0} \tag{11}$$

Based on Fick's law analogy, the PST described in (10) closely resembles the one proposed by Lind et al. [13]. It has been widely utilized in recent years within SPH for its effectiveness in preventing tensile instability issues and improving simulation results. While PST models in SPH conventionally operate within the confines of virtual particle layers, the research conducted by [28] extended the methodology of [13] for particle shifting to boundary integrals. However, applying the PST directly for particle initialization, particularly near the boundary, is not feasible, as will be discussed shortly.

First, to calculate the particle shift $\delta x_a$ in the BIPI algorithm, we employ the following approach: Since equations (10) and (11) reduce to the diffusion law for PST, as discussed by Lind et al.[13], we set $D_a = \mathcal{J}\,h^2$. This value is derived from thermodynamic considerations of $p_0$ to be proportional to $\rho_0$, and by imposing a CFL constraint on the time step $\delta t$. Next, to prevent extensive shifting of particles, we establish an upper limit of $|\delta x_a| = 0.5\,dx_r$, which is half the average particle spacing. In this investigation, we use $\mathcal{J} = 0.5$ unless otherwise specified.

Thus, reducing our particle shift equation (10) to,

$$\delta x_a = \begin{cases} \widetilde{\delta x}_a & |\widetilde{\delta x}_a| < 0.5\,dx_r, \\ -0.5\,dx_r\,\dfrac{\nabla C_a}{\|\nabla C_a\|} & 0.5\,dx_r < |\widetilde{\delta x}_a| \end{cases} \tag{12}$$

While using (12) directly can redistribute particles to minimize $\nabla C_a$, it can inadvertently result in particles being positioned too close to the boundary. However, the layer of particles adjacent to the boundary must remain uncompressed. Assuming a particle's volume is uniformly distributed in the radial direction, with a particle radius of $\frac{dx_r}{2}$, maintaining a uniform particle distribution dictates that the particle near the boundary should be positioned $\frac{dx_r}{2}$ away from the boundary. This issue is typically addressed in physics-driven simulations by applying external forces by the boundary. For example, in flow simulations, the boundary pressure term, $f_s = p_s$ in (8), increases as particles approach the boundary. While various methods for updating $p_s$ have been explored in recent literature for fluid flows [21], implementing such techniques directly for BIPI would necessitate solving the complete set of fluid equations, thereby undermining the purpose of particle initialization schemes.

The acceleration due to the pressure gradient can be expressed in terms of concentration gradient and varying boundary pressure as,

$$\ddot{x}_a = -\frac{2p_a}{\rho_a}\left[\nabla C_a - \frac{1}{\gamma_a}\sum_s \frac{1}{2}\left(\frac{p_s}{p_a}-1\right)\nabla\gamma_{as}\right] \quad (13)$$

Which gives the particle displacement term,

$$\widehat{\delta x_a} = -D_a\left[\nabla C_a - \frac{1}{\gamma_a}\sum_s \frac{1}{2}\left(\frac{p_s}{p_a}-1\right)\nabla\gamma_{as}\right] \quad (14)$$

Accurately modeling $p_s$ proves challenging and relies on the method used to model volume compression. For instance, if we employ small non-linear springs at the boundary to simulate particle compression and apply conservative force, as shown in Figure 3a, the pressure variation becomes non-linear (Figure 3b). However, our numerical experiments have shown that a simple step function ensures particles remain approximately a radius away from the boundary (see Figure 3c). We opt for $p_s = \begin{cases} p_a & |(x_a - x_s)\cdot\hat{\underline{n}}_s| \geq 0.5\,dx_r \\ 2\,p_a & |(x_a - x_s)\cdot\hat{\underline{n}}_s| < 0.5\,dx_r \end{cases}$ in equation (14).

Applying an upper bound to the total particle shift, as before, we derive:

$$\delta x_a = \begin{cases} \widehat{\delta x_a} & |\widehat{\delta x_a}| < 0.5\, dx_r, \\ 0.5\, dx_r \dfrac{\widehat{\delta x_a}}{\|\widehat{\delta x_a}\|} & 0.5\, dx_r < |\widehat{\delta x_a}| \end{cases} \quad (15)$$

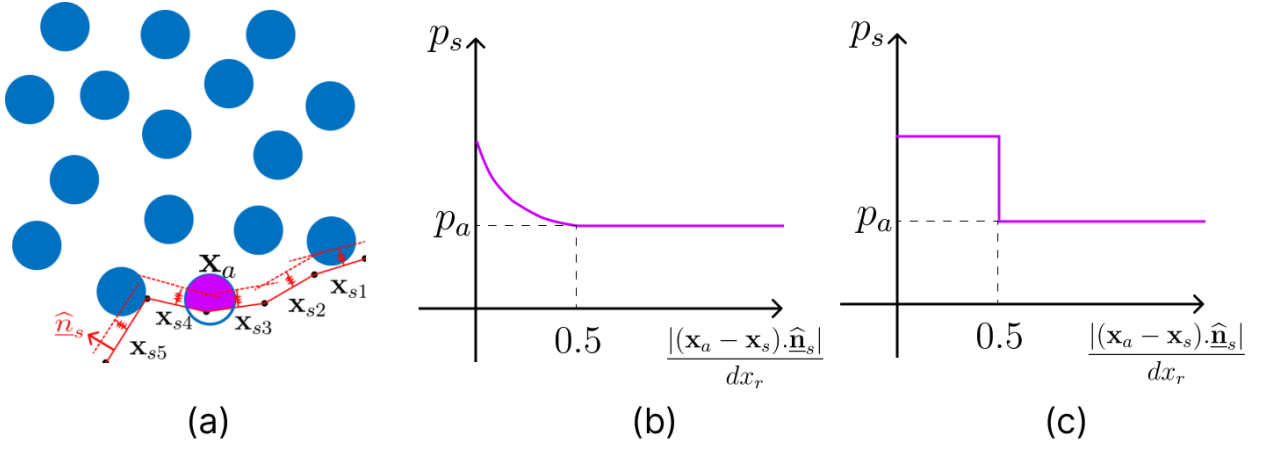

*Figure 3 Modeling external repulsive force for particles immediately adjacent to the boundary. (a) Illustration of force exerted by particles on neighboring particles within their radius, (b) representation of non-linear boundary pressure variation, and (c) representation of a step force acting on particles near the boundary. Here $p_s$ is the pressure at the boundary location $x_s$, $p_a$ is the pressure at location $x_a$, $dx_r$ is the particle spacing, and $\widehat{\mathbf{n}}_s$ is the inward surface normal.*

Furthermore, we modify our algorithm to expedite settling time and minimize the computational overhead. This involves initially shifting particles with truncated kernels and subsequently shifting particles everywhere except those immediately adjacent to the boundary. This final adjustment works best when particles are generated on a background Cartesian grid. Otherwise, we recommend iteratively employing (15) in Step 2 until achieving the desired uniform configuration, when $\nabla C_a$ is minimized.

To define the stopping criteria in the algorithm, we use two parameters, $TPD_{avg}$ and $|\nabla C|_{avg}$, which are defined as $TPD_{avg} = \frac{1}{n_{pack}} \sum_{a=1}^{n_{pack}} TPD_a$ and $|\nabla C|_{avg} = \frac{1}{n_{pack}} \sum_{a=1}^{n_{pack}} |\nabla C_a|$, where $n_{pack}$ is the number of packable particles, indicating only those particles that are included in the packing algorithm step. Here the Total Particle Displacement (TPD) for each particle is defined as the change in particle position from its starting position, that is, $|x_{a_i} - x_{a_0}|$, where $i$ is the iteration number in Step2.

As starting on the Cartesian grid is generally the most practical approach, we outline our final modification as follows:

### 3.2.1 Step 2a: Pack particles adjacent to the boundary

We begin by identifying particles located at a distance $k_a = \kappa h$ from the boundary as 'packable particles,' ensuring that all particles with truncated kernels are included. This step allows us to focus only on a small layer of particles (shown in green in Figure 4a) and their potential neighbors (shown in blue in Figure 4a), collectively referred to as 'selected particles.' These 'selected particles' are identified during the first iteration of Step 2a, and in all subsequent iterations, the nearest neighbor algorithm for identifying particle pairs is applied exclusively to these 'selected particles.'

This process significantly improves the computational efficiency of particle packing, as it is unnecessary to consider internal particles until those adjacent to the boundary align with the boundary shape. The reasoning behind this is that boundary reshaping first affects particles immediately adjacent to the boundary, and these disturbances gradually travel inward to the interior particles. Unless the boundary-adjacent particles stop moving, these disturbances will continue to propagate, leading to unnecessary updates in the positions of internal particles from the start of the simulation.

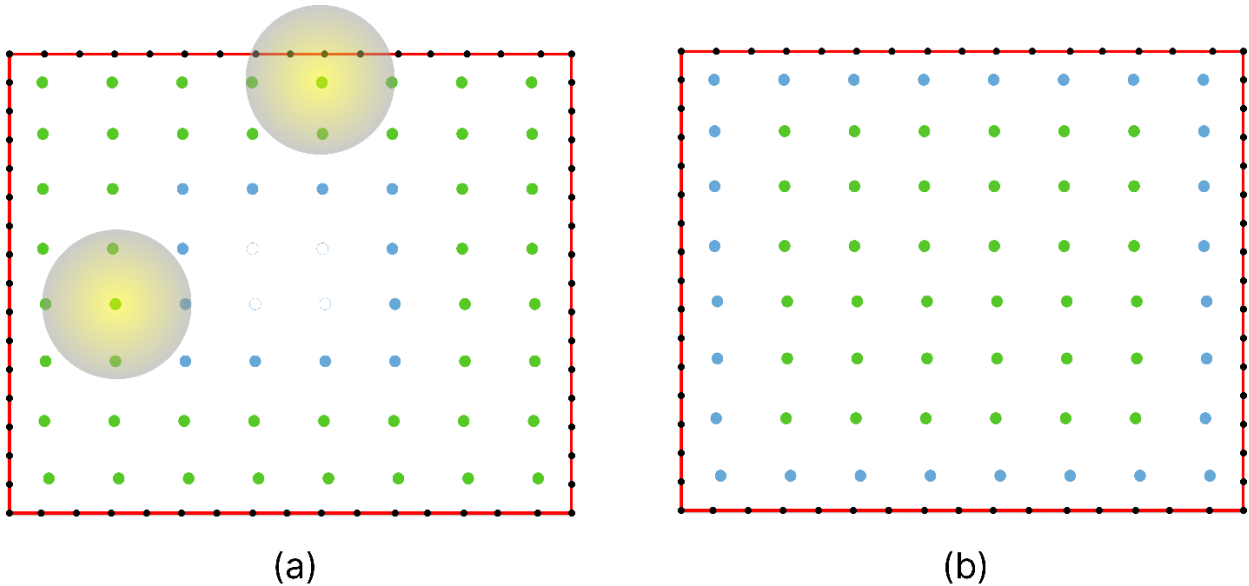

*Figure 4 Particles and boundary elements in consideration for (a) in Step 2a and (b) in Step 2c. Here, the packable particles shown in green are shifted using (15) in an iteration, whereas the blue particles are only used for kernel support.*

By restricting the algorithm to a small layer of 'selected particles' with a width of approximately $k_a + \kappa h$, we significantly reduce computational load. Although our packing algorithm employs (15), which aims to minimize $|\nabla C|_{avg}$, we found that $TPD_{avg}$ is a better parameter for determining the stopping criteria in Step 2a. The goal in this step is to allow particles next to the boundary to conform to the boundary shape. In our numerical experiments, we observed that $TPD_{avg}$ initially increases as particles settle and then stabilizes, indicating particle settlement. Therefore, it is sufficient to stop Step 2a after a desired number of iterations or when the change in $TPD_{avg}$ is less than 1%.

### 3.2.2 Step 2b: Freeze particle layer adjacent to the boundary

Next, we freeze the positions of particles that are immediately adjacent to the boundary. These are particles located within a distance $k_b$ from the boundary, where $k_a >> k_b > 0.5 dx_r$. The lower limit for $k_b$ ensures that at least the first layer of boundary particles is frozen, while the upper limit allows for further minimization of $\nabla C_a$ for the packable particles in Step 2a that form the transition layer from boundary-adjacent particles to interior particles. For the Wendland kernel tested in this investigation, $k_b = 0.6 dx_r$ was found to be a suitable value that worked well across various complex shapes and $h/dx_r$ ratios. In our code implementation, this is achieved by identifying particles whose $\gamma_a$ values are less than the $\gamma_a$ value of a particle located at a distance $k_b$ from a straight line, and then freezing these particles in place.

### 3.2.3 Step 2c: Pack internal particles

Finally, we update positions of all the particles which are not frozen in Step2b to obtain a configuration with minimal $|\nabla C|_{avg}$. All particles in the domain are now "selected", but only the particles unfrozen in Step2b are considered "packable". Step 2c is stopped when $|\nabla C|_{avg}$ of the particles varies less than 1% or when desired number of iterations of Step2c is hit. Convergence of $|\nabla C|_{avg}$ is not necessarily observed, as particle re-arrangement continues to happen since (6) and (7) cannot be achieved in practice. However, our objective is to use configuration whose order of $|\nabla C|_{avg}$ is low. From our experience, Step 2c typically requires only several hundred iterations unless $\nabla C_{avg}$ begins to increase, which would indicate a transition to the next conditionally stable configuration. As long as $\nabla C_{avg}$ does not increase significantly in its order of magnitude, Step 2c can be safely terminated. It's also important to note that the concentration gradient for the particle arrangement depends on the smoothing length-to-particle spacing ratio ($h/dx_r$) and the value of the smoothing length ($h$), meaning that the number of iterations required to reach particle configuration suitable as the starting configuration, can vary.

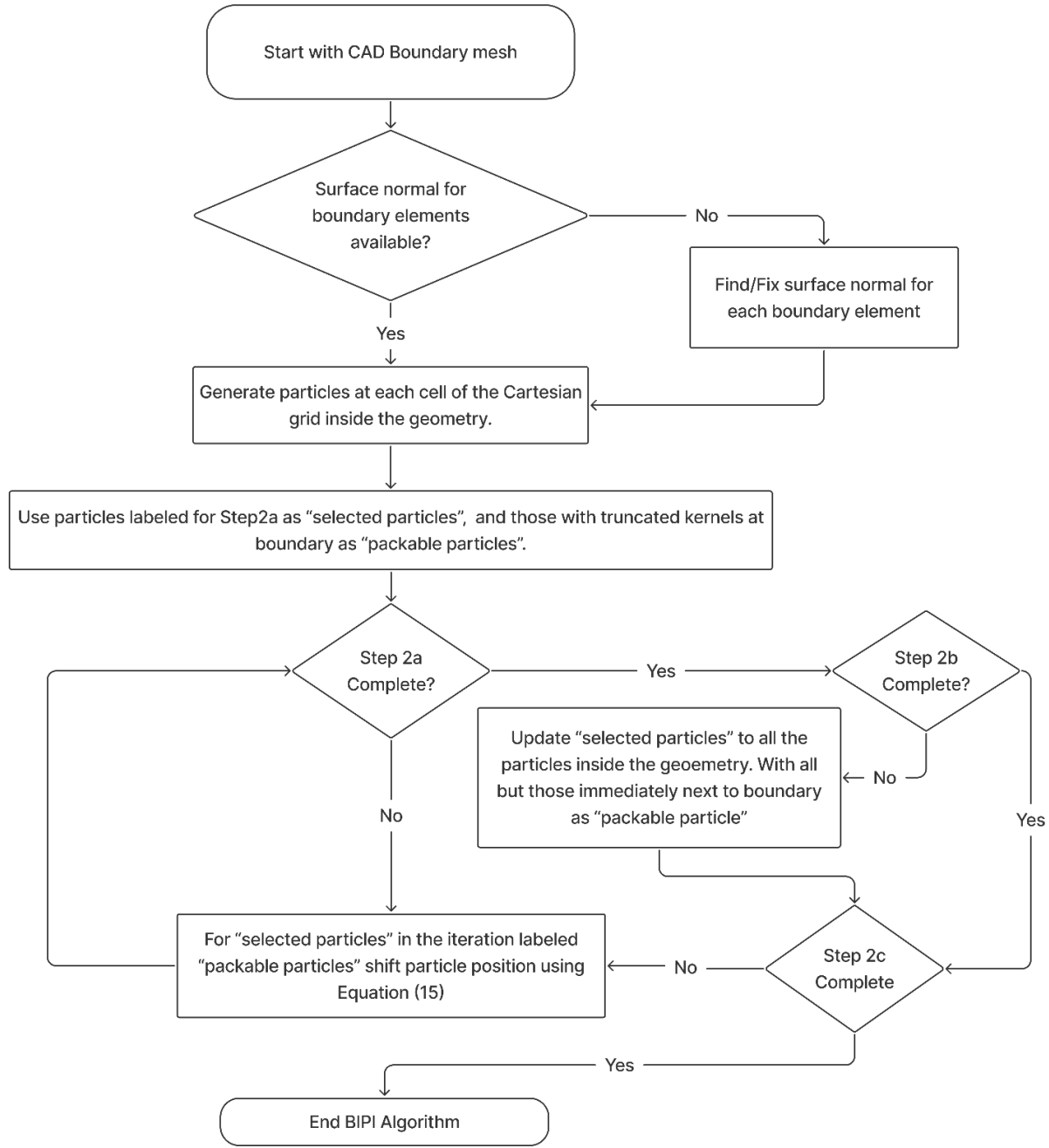

*Figure 5 BIPI algorithm flow chart for code implementation (code available on https://github.com/RealTimeSimLab/BIPI2D.git ).*

The algorithmic steps of the BIPI algorithm is shown in the flowchart in Figure 5, and the software implementation of the algorithm along with sample scripts to generate particles is made available on GitHub repository "https://github.com/RealTimeSimLab/BIPI2D.git".

# 4  Results and Discussion

## 4.1  BIPI algorithm on different geometries

To illustrate the BIPI algorithm discussed in the preceding section, we examine the trapezoidal geometry depicted in Figure 2. Initially, the geometry's interior is discretized into particles using the underlying Cartesian grid, giving us the starting configuration for Step 2. We closely examine the particle arrangement at the top-left and top-right corners of the trapezoidal geometry, as shown in Figure 6, to highlight the significance of employing a force-based equation (15), instead of equation (12).

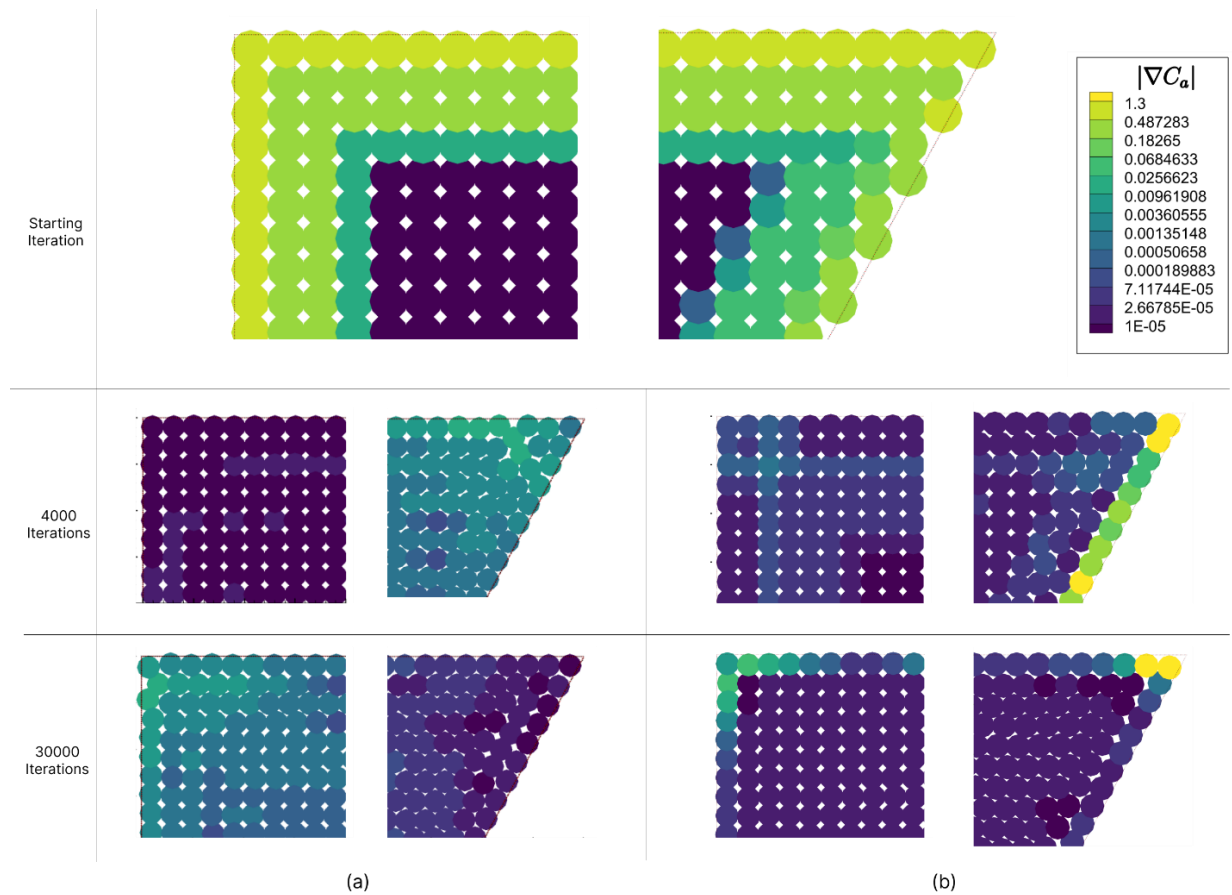

*Figure 6 Comparison of particle packing at 4000 iterations and 30000 iterations using (a) equation (12), (b) equation (15) with Step2a run for first 7000 iterations Here $|\nabla C_a|$ is the concentration gradient of each particle.*

It is seen that using equation (12) for the BIPI algorithm causes particles to approach the boundary too closely while others move away from it, showing that minimizing $|\nabla C_a|$ alone is insufficient to maintain a uniform configuration near the boundary. This observation is confirmed by the $|\nabla C_a|$ calculation for the starting configuration at the top-left corner, where, despite the particles following the boundary precisely, the $|\nabla C_a|$ value for truncated kernels remains non-zero. Additionally, employing equation (12) presents challenges in establishing a suitable stopping criterion for the BIPI algorithm. However, using equation (15) shows that the BIPI algorithm effectively enables particle redistribution along the boundary while also reducing $|\nabla C|_{avg}$. By including an external force in the formulation, particles near the boundary maintain a consistent distance from it, approximately one particle radius away. Additionally, setting stopping criteria using $TPD_{avg}$ for Step 2a and $|\nabla C|_{avg}$ for Step 2c is straightforward following the procedure highlighted in Section 3.2.

To understand the stopping criteria defined in Step 2a and Step 2c, we use (15) to pack particles in the trapezoidal geometry for various particle spacing configurations using $h/dx_r = 2$. First, we run Step 2a for 20,000 steps and Step 2c for 40,000 steps. In Figure 7, solid lines represent Step 2a, while dotted lines represent Step 2c. The $TPD_{avg}$ curve in Figure 7a shows that $TPD_{avg}$ stabilizes after several thousand iterations, as indicated by the solid lines. However, while $|\nabla C|_{avg}$ initially decreases, it does not provide a clear stopping criterion, as indicated in Figure 7b. This is because the error in $\nabla C_a$ is dominated by particles immediately next to the boundary, where the concentration gradient remains high due to kernel truncation and the consistency of the gradient formulation used to calculate $\nabla C_a$. Therefore, $TPD_{avg}$ serves as a better indicator for stopping Step 2a.

In contrast, once Step 2c begins, we exclude particles next to the boundary from the calculation of $|\nabla C|_{avg}$, which allows us to better capture its reduction during the simulation. The dotted lines in Figure 7b show that $|\nabla C|_{avg}$ decreases significantly within the first few iterations. As particles continue to rearrange, further reduction of $|\nabla C|_{avg}$ occurs. However, after dropping several orders of magnitude, continuous decline is not always observed. For example, in the dotted line for $dx_r = 0.04$, $|\nabla C|_{avg}$ increases slightly around iteration 45,000 before dropping again, and for $dx_r = 0.02$, a similar slight increase occurs around iteration 35,000 before declining again. A similar behavior is expected for $dx_r = 0.01$ at a higher iteration count. Importantly, these slight increases in $|\nabla C|_{avg}$ remain within the same order of magnitude and simply indicate that configurations tend to rearrange since $\nabla C_a$ for all particles is never zero at any given instant. Despite this, their values become small enough to allow us to terminate Step 2c earlier.

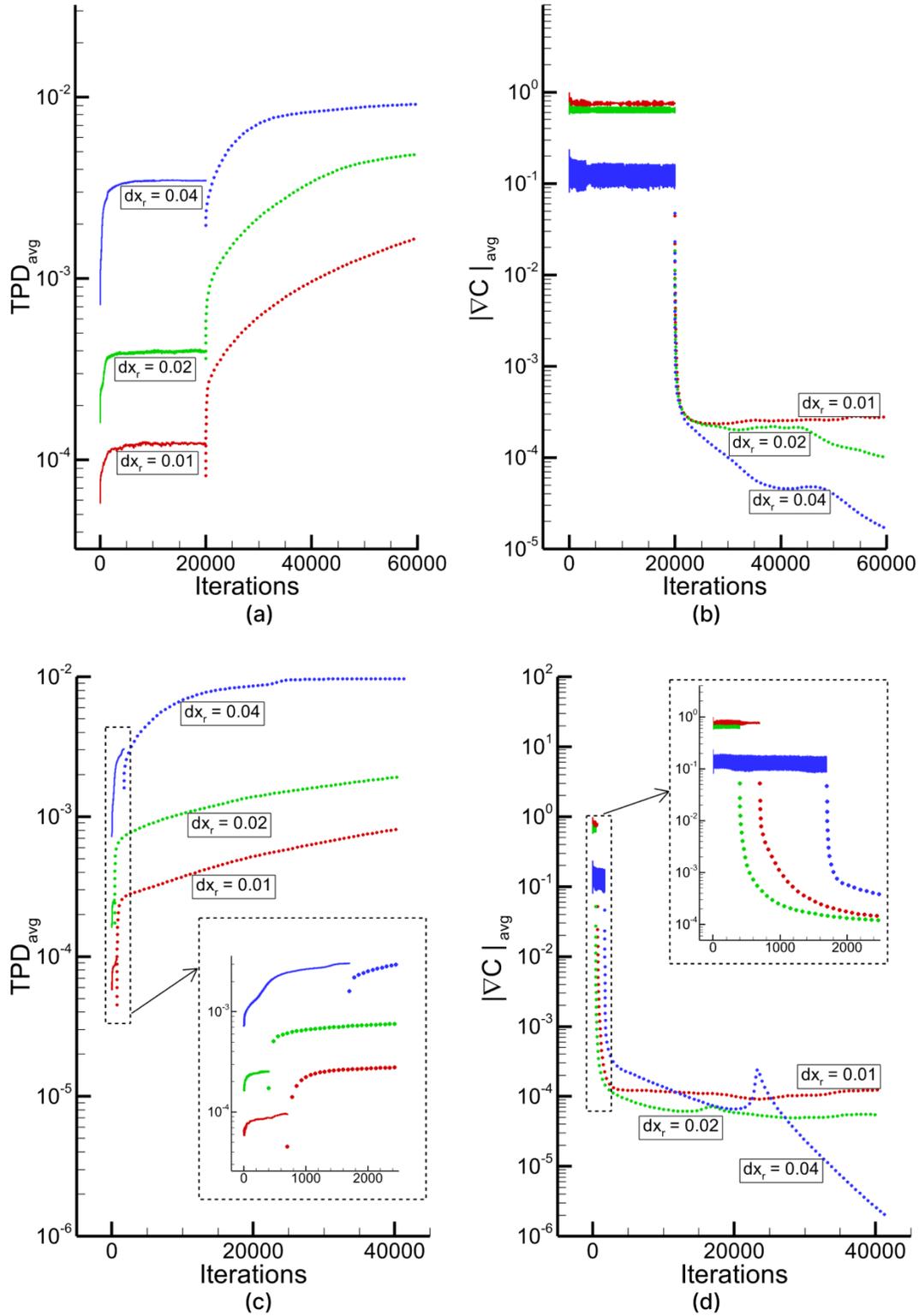

*Figure 7 Particle packing for particle spacing $dx_r = 0.02m$ in the trapezoidal geometry showing the evolution of: (a) $TPD_{avg}$ and (b) $|\nabla C|_{avg}$ for 20,000 iterations in Step 2a and 40,000 in Step 2b; (c) $TPD_{avg}$ and (d) $|\nabla C|_{avg}$ using 1% $TPD_{avg}$ as the stopping criterion for Step 2a.*

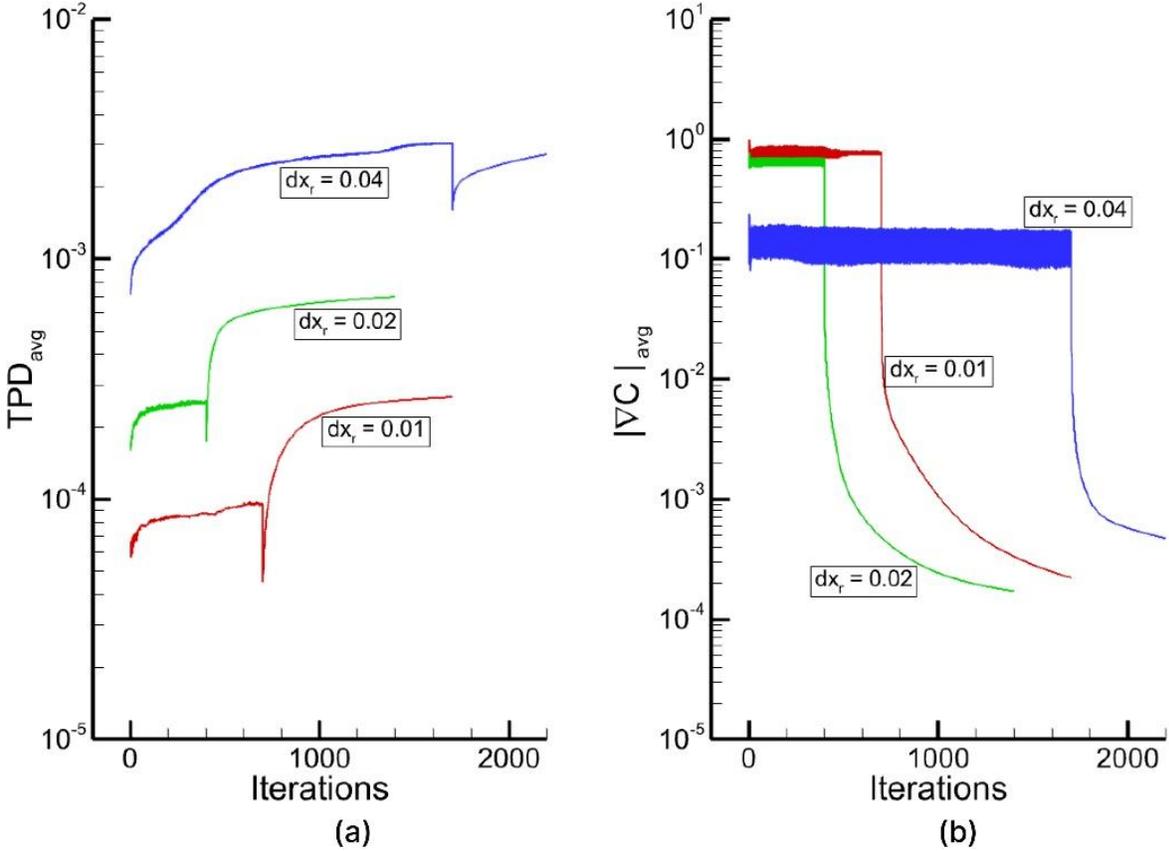

*Figure 8 Evolution of (a) $TPD_{avg}$ and (b) $|\nabla C|_{avg}$ during the BIPI algorithm, for particle spacing $dx_r = 0.01m$, $dx_r = 0.02m$, and $dx_r = 0.04m$*

This is precisely why $TPD_{avg}$ is not used as an indicator for Step 2c; changes in particle positions and the emergence of new configurations with different 'local minima' of $|\nabla C|_{avg}$ would invariably lead to increases in $TPD_{avg}$. In practical simulations where damping parameters such as viscosity or artificial viscosity are used, small errors in $|\nabla C|_{avg}$ do not significantly affect the simulation results. If there is a need to improve the gradient calculation itself, other techniques, such as improving the consistency of the operator or using a smoothing kernel with a higher order of convergence, should be considered.

Next, we run Step2a till $TPD_{avg}$ change is within 1% and run Step2c for 40000 iterations. The $TPD_{avg}$ graph of Step2a region in the magnified region of Figure 7c and Figure 7d indicates that significantly less iterations are required to stop Step2a. The Step2c section of Figure 7d again indicate that $|\nabla C|_{avg}$ initially drops significantly by several orders of magnitude. Like in Figure 7a and Figure 7b, rearrangements are expected to occur, but these are still several orders smaller than initial errors in $|\nabla C|_{avg}$.

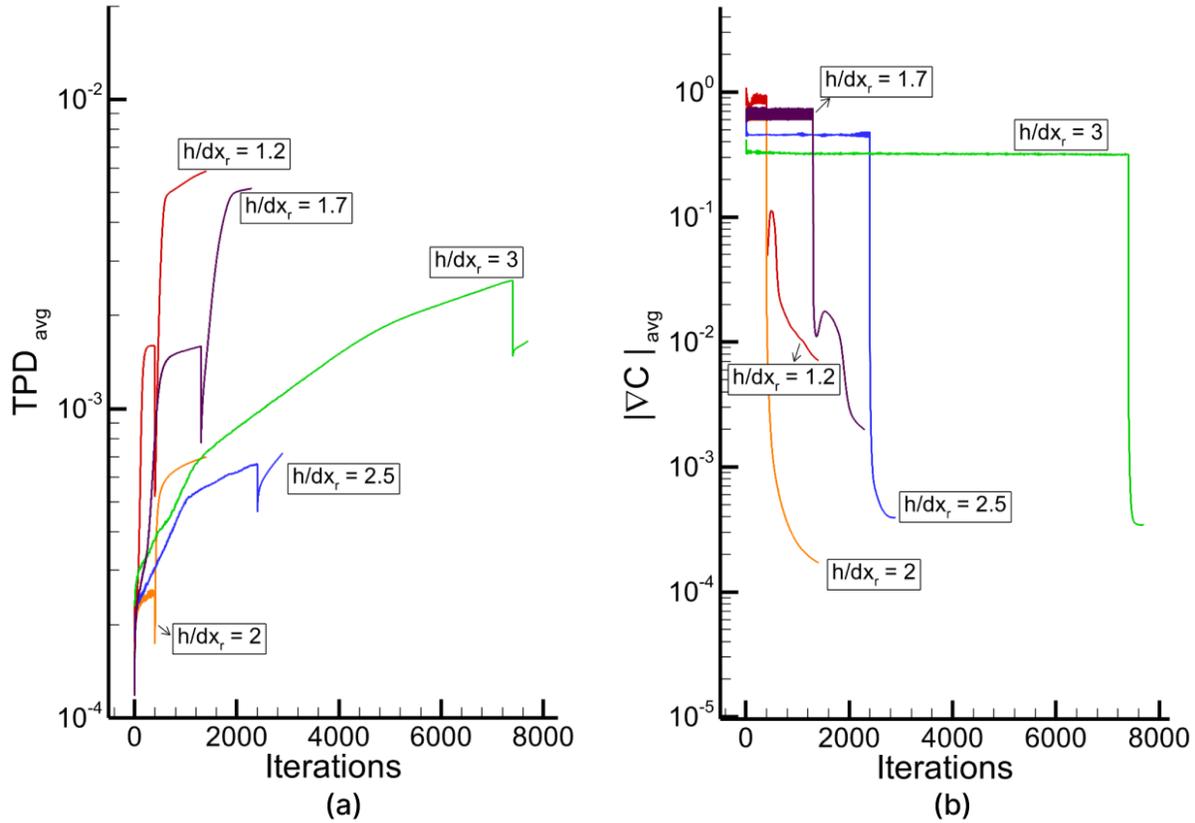

Figure 9 Evolution of (a) $TPD_{avg}$, (b) $|\nabla C|_{avg}$, during the BIPI algorithm for various smoothing length to particle spacing ($h/dx_r$) ratios for the Trapezoid geometry.

Now, to demonstrate the effectiveness of the algorithm across different particle spacings and smoothing lengths, we present Figures 8 and 9. In Figure 8, the algorithm is applied for $h/dx_r = 2$ with $dx_r = [0.01, 0.02, 0.04]$, and in Figure 9, for $h/dx_r = [1.2, 1.7, 2, 2.5, 3]$ with $dx_r = 0.02$. Unlike in Figure 7, the BIPI algorithm implemented in Figure 8 and Figure 9 is stopped at 1% change of $TPD_{avg}$ in Step 2a and 1% change of $|\nabla C_{avg}|$ in Step 2c. It is observed that more iterations of the BIPI algorithm are generally required to achieve a packed configuration as the smoothing length increases for a given particle spacing. This is because, with a higher $h/dx_r$ ratio, more particles have kernels truncated at the boundary, necessitating additional iterations in Step 2a for redistribution. However, this is not always the case, as seen with the run for $h/dx_r = 1.7$, which requires more iterations than for $h/dx_r = 2$. Therefore, while the number of iterations in Step 2a and Step 2c is influenced by the smoothing length ratio and particle spacing, its behavior ultimately depends on the kernel values in the discrete approximation. In fact, maintaining a regular particle arrangement, along with an appropriate choice of $h/dx_r$, is crucial for achieving accurate results in SPH. For instance, Price's work [6] demonstrates that the error in discrete kernel gradients is minimized for certain smoothing lengths. While Price's study focused on a 1D kernel, this property holds true for kernels in 2D and 3D problems as well. Our results also reflect this; a more significant factor influencing the number of iterations is the choice of kernel and the corresponding smoothing length.

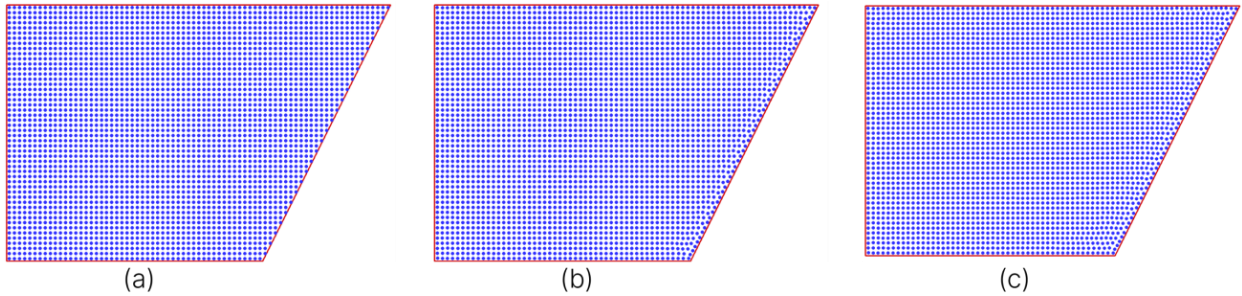

*Figure 10 Scatter plot (a) at the start of BIPI algorithm, (b) after 400 iterations of Step 2a stopped when $TPD_{avg}$ is converged within 1% change, (c) after 1000 iterations of Step 2c stopped when $\nabla C_{avg}$ is converged within 1% change.*

In Figure 7-9, it is also noticed that $TPD_{avg}$ suddenly increases after the end of Step 2a, because of the larger concentration gradient developed for interior particles adjacent to "packable particles". The dip and then increase is because of how $TPD_{avg}$ is calculated with regards to $n_{pack}$. Since, from Step 2a to Step 2c, we consider more particles in the domain, which have not yet moved from their starting position, we notice a dip in the $TPD_{avg}$ value. The concentration gradient leads to sudden redistribution of those particles which are previously unmoved in Step2a, to minimize their concentration gradient, leading to increased $TPD_{avg}$ value. This can be seen in the particle plot of Figure 10, where the BIPI algorithm with the stopping criteria discussed in Section 3.2 is used for particle spacing of 0.02, corresponding to the green line in Figure 8.

Similarly, in Step 2c, $|\nabla C_{avg}|$ does not include the particles immediately adjacent to the boundary, which hold the higher concentration gradient as they are kept half the particle spacing away from the boundary in Step2a using (15). Hence the sudden change in values in both $|\nabla C_{avg}|$ and $TPD_{avg}$ in Step 2c is simply because of our choice in determining the parameter for BIPI algorithm and how it is calculated mathematically.

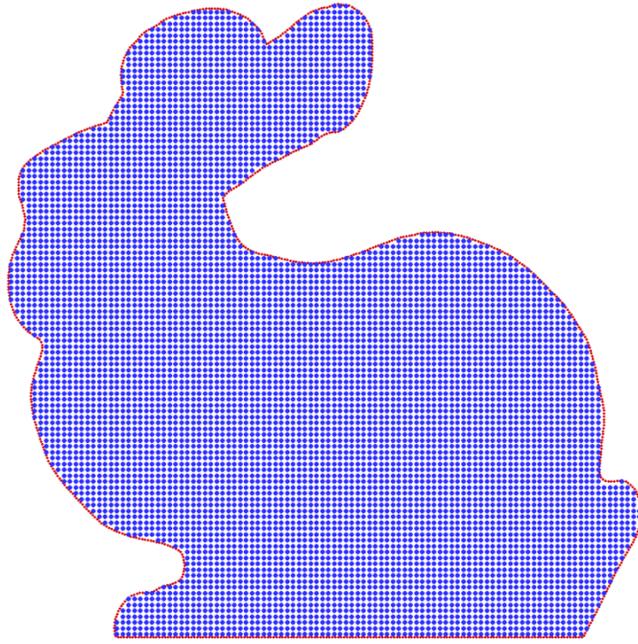

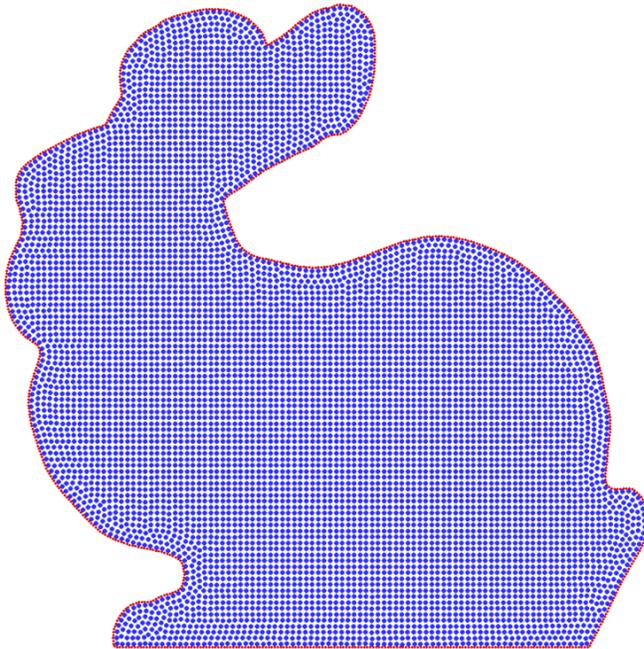

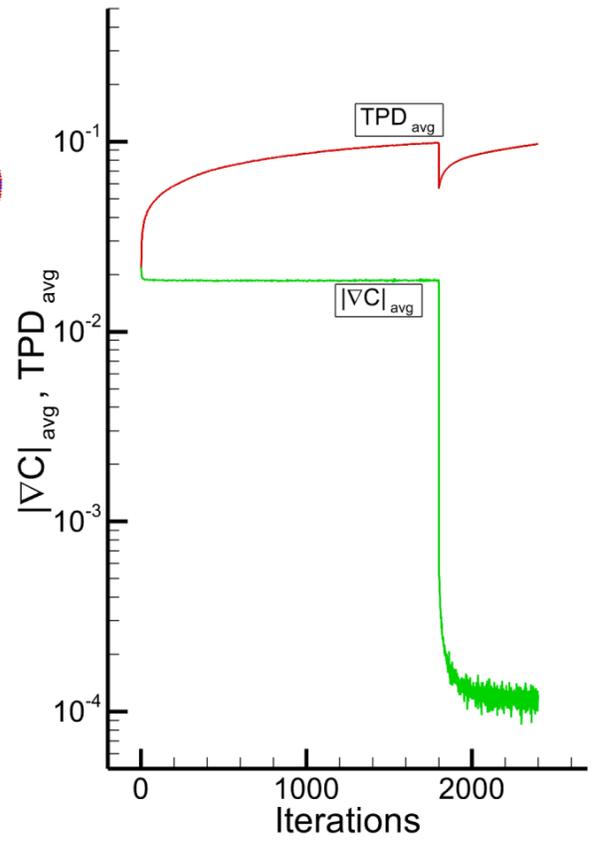

*Figure 11 BIPI algorithm on Standford 2D Bunny showing (a) Starting Configuration, (b) Initialized particle distribution, (c) $TPD_{avg}$ and $|\nabla C|_{avg}$ evolution with iterations*

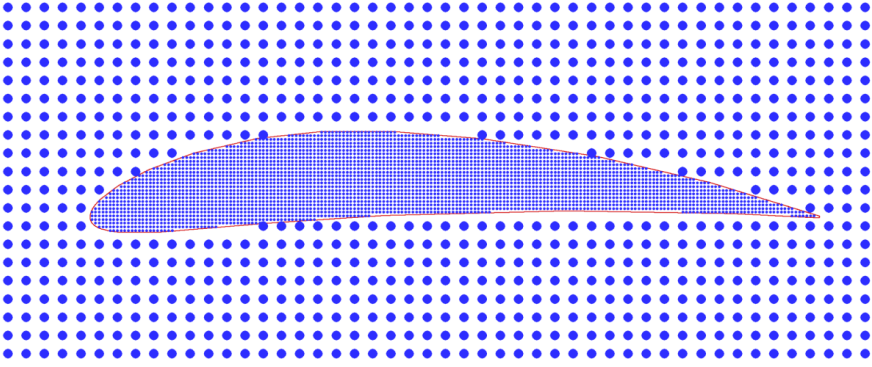

(a)

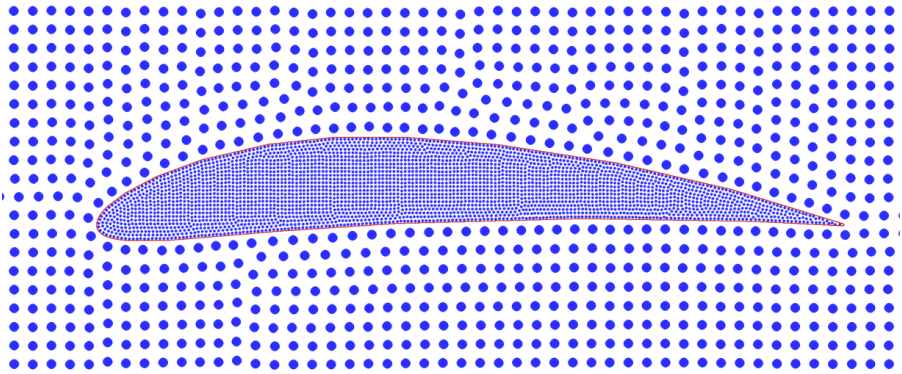

(b)

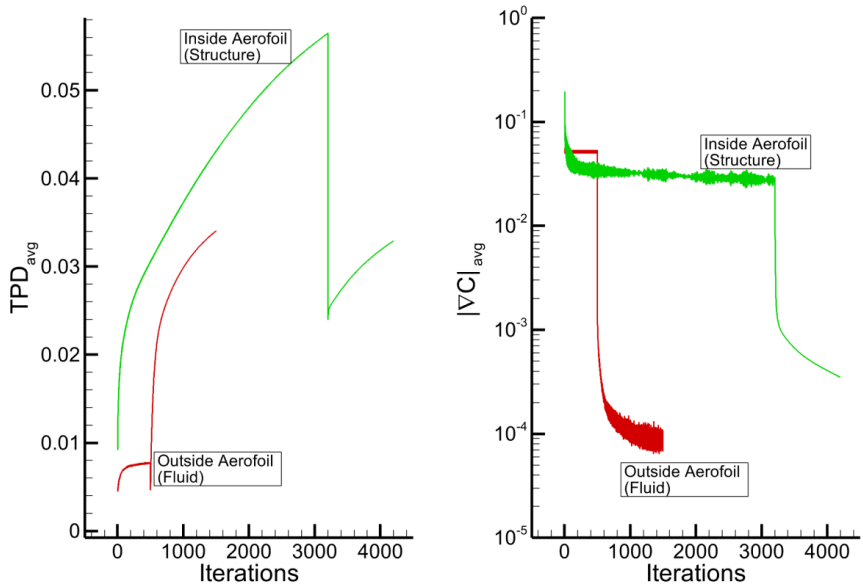

(c)

*Figure 12 Application of the BIPI algorithm to a NACA6412 Airfoil with varying particle spacing for the airfoil structure and fluid domain, with plots showing: (a) Starting Configuration, (b) Initialized particle distribution, (c) $TPD_{avg}$ and $|\nabla C|_{avg}$ evolution with iterations.*

We also show the algorithm's performance on two additional complex geometries: a 2D Bunny in Figure 11 and a NACA airfoil in Figure 12. Again, in both these graphs, we observe the sudden change in $|\nabla C|_{avg}$ and $TPD_{avg}$ curves due to the change in particles considered "packable particles" in BIPI algorithm. In both cases, we notice that the particles follow the boundary of the complex shapes while also reducing the average concentration gradient when the BIPI algorithm is stopped.

### 4.2 BIPI for fluid simulations

#### 4.2.1 Hydrostatic Simulations

Since SPH is widely used in fluid flow problems, we test the effect of particle packing on fluid dynamics simulations. These examples also illustrate when the benefits of using a packing algorithm are most prominent. One of the objectives of this work is to implement the BIPI algorithm in BISPH research. We examine the BISPH equations governing a weakly compressible flow problem, showing the transition from an initial particle configuration to a packed particle configuration for flow simulation without using virtual particle layers. The weakly compressible equations used in this section are given by:

$$\frac{D\rho_a}{Dt} = -\rho \frac{1}{\gamma_a} \left( \sum_b (\boldsymbol{v}_b - \boldsymbol{v}_a) \boldsymbol{\nabla}_a W \, V_b - \sum_s (\boldsymbol{v}_s - \boldsymbol{v}_a) \boldsymbol{\nabla} \gamma_{as} \right) + \mathcal{D}_a \qquad (16)$$

$$\frac{D\boldsymbol{v}_a}{Dt} = -\frac{1}{\rho_a \gamma_a} \left( \sum_b (p_a + p_b) \boldsymbol{\nabla}_a W \, V_b - \sum_s (p_a + p_s) \boldsymbol{\nabla} \gamma_{as} \right) + \frac{\mu}{\rho_a} \boldsymbol{\nabla} \cdot \boldsymbol{\nabla} \boldsymbol{v}_a + \boldsymbol{F}_{ext} \qquad (17)$$

$$p_a = \frac{\rho_0 c_0^2}{7} \left( \left( \frac{\rho_a}{\rho_0} \right)^7 - 1 \right) + P_B \qquad (18)$$

In equations (16) to (18), $\frac{D}{Dt}$ denotes the total derivative with respect to time, $\boldsymbol{v}$ represents velocity, $\mathcal{D}_a$ is the density diffusion term, $\mu$ represents fluid viscosity, $\boldsymbol{F}_{ext}$ denotes the external force acting per unit mass, and $c_0$ represents the speed of sound. Additionally, $P_B$ in (18) is the background pressure which is 0 for free surface flows and a positive value in confined flow simulation. The density diffusion term is used to reduce density noise in weakly compressible SPH, as suggested in [32]. The speed of sound $c_0$ is chosen such that the Mach number is less than 0.1, allowing for density changes to remain within 1%, to justify the weakly compressible assumption. In all our simulations, we solve (16)-(18) using a simple explicit scheme (using first order Euler scheme), with the time step $\delta t$ selected using the criteria:

$$\delta t = \min\left(0.25\frac{h}{c_0}, 0.25\sqrt{\frac{h}{|\boldsymbol{F}_{ext}|}}, 0.125\, h^2,\ 0.125\frac{\rho_0 h^2}{\mu}\right) \quad (19)$$

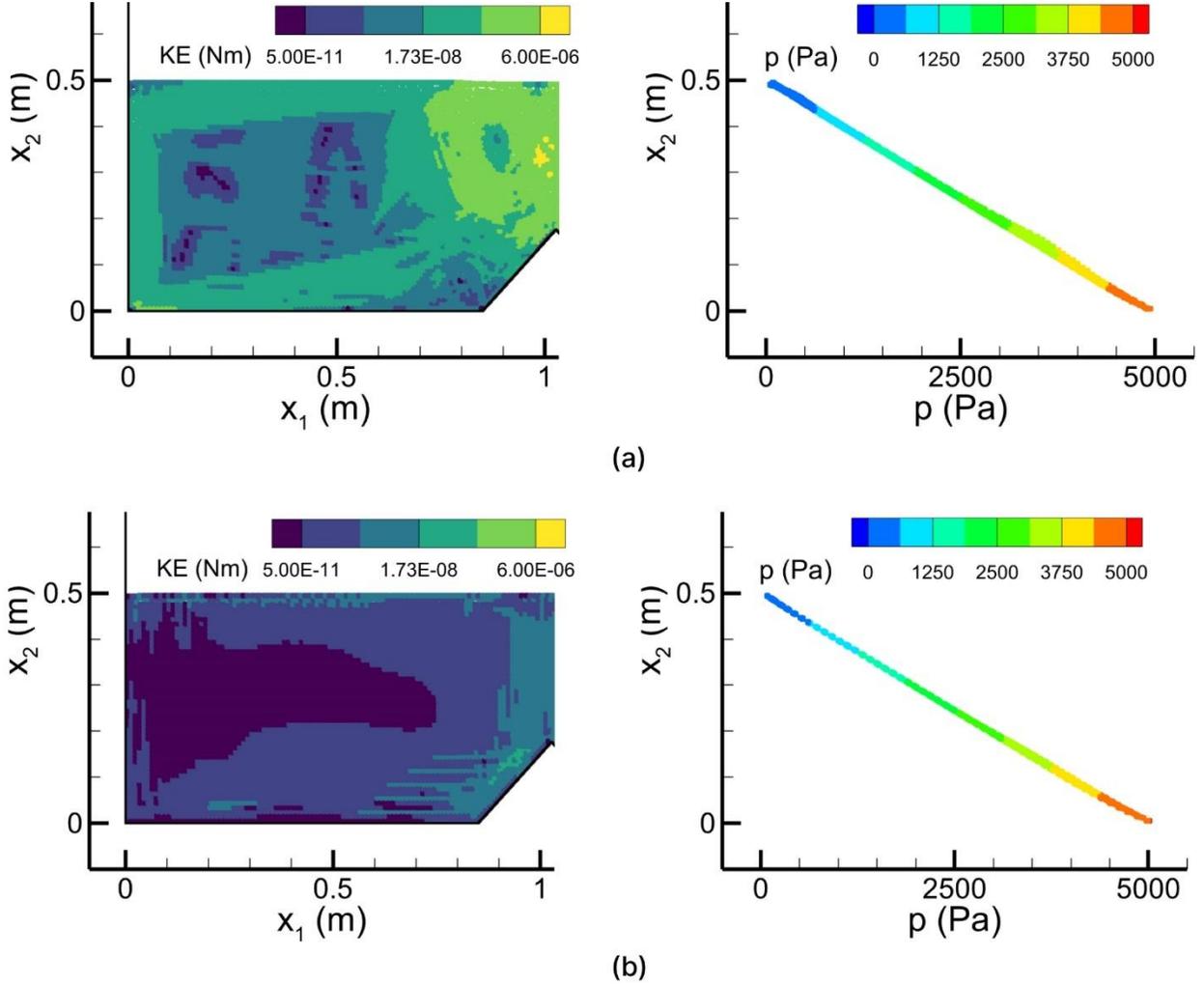

Figure 13 Hydrostatic simulation of wedge in a tank after 20s, for particle spacing $dx_r = 0.01m$ and smoothing length $h = 0.02m$, for (a) Without BIPI pre-processing, (b) With BIPI preprocessing. Here KE is the Kinetic Energy, p is the pressure, $x_1$ is the horizontal coordinate and $x_2$ is the vertical coordinate.

Further, the boundary pressure is modeled as:

$$p_s = p(\tilde{\rho}_a) + \rho_a c_0 (\boldsymbol{v}_s - \boldsymbol{v}_a) \cdot \widehat{\underline{\boldsymbol{n}}}_s \quad (20)$$

where $p(\tilde{\rho}_a)$ is calculated using (18) to account for fluid compression near the boundary. The compressed density $\tilde{\rho}_a$, adjusted for particles approaching the boundary, is defined by:

$$\tilde{\rho}_a = \begin{cases} \rho_a & |x_a - x_s| > 0.5\, dx_r \\ \dfrac{2\,\rho_a}{dx_r}(dx_r - |x_a - x_s|) & |x_a - x_s| < 0.5\, dx_r \end{cases} \quad (21)$$

A key test for the stability and accuracy of SPH formulations is the hydrostatic problem, where a linear pressure variation in the fluid can be verified for various geometries. In these simulations, we set $F_{ext_2} = -9.81\, m/s^2$ (to model acceleration due to gravity along water depth) and $\rho_0 = 1000\, Kg/m^3$. Since no artificial viscosity term is used in (17), a higher viscosity of $\mu = 10\, Pa\, s$ is applied to dampen noise, as suggested in [33] for hydrostatic simulations.

We test two hydrostatic tanks: a simple rectangular tank with a wedge and a more complex trapezoidal tank with coral. For each, hydrostatic pressure is set up at $t = 0s$ and the simulation is run for an extended time by solving (16)-(18). Simulation accuracy and stability are evaluated by verifying the linearity of the pressure variation and the near-zero kinetic energy (KE) of the system.

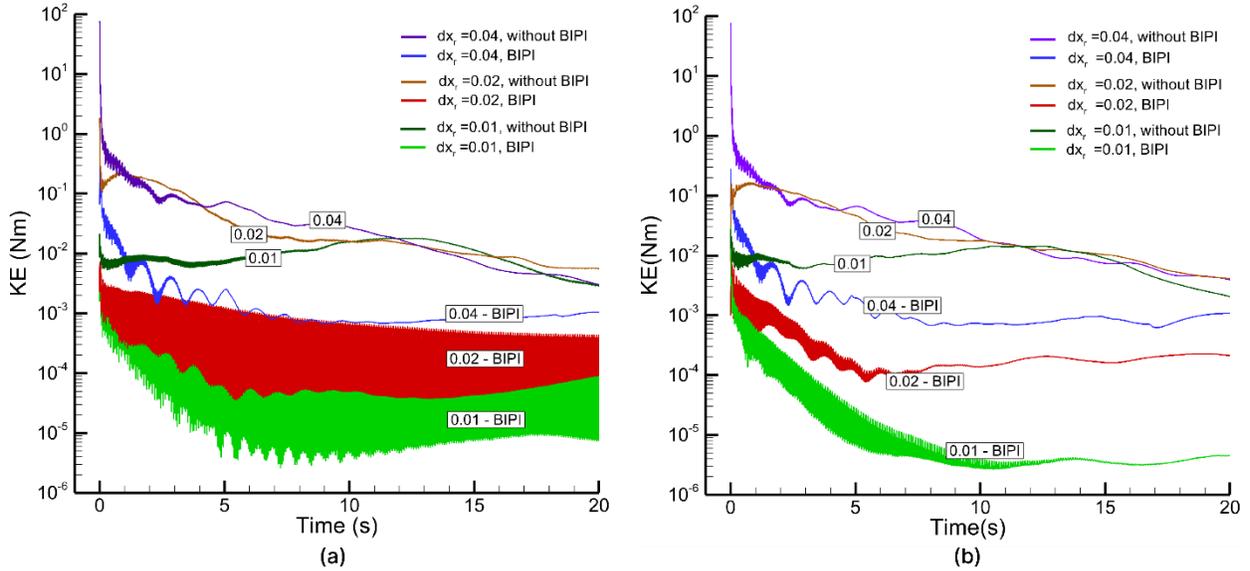

*Figure 14 Comparison of Kinetic Energy (KE) evolution using different particle spacing($dx_r$) in hydrostatic of tank with wedge simulation, with and without BIPI preprocessing step, using (a) boundary layer theory formulation and (b) particle consistent gradient formulation, evaluating the viscous term $\dfrac{\mu}{\rho_a} \nabla \cdot \nabla v_a$.*

The rectangular tank with a wedge has a width of 2.05 m and a wedge height of $\sqrt{2}/8$ m, as used in [33]. Water fills the tank up to 0.5 m, and particles are initialized on a Cartesian grid. The viscous term $\dfrac{\mu}{\rho_a} \nabla \cdot \nabla v_a$ is modeled using the boundary layer theory from [21,28]. Running the hydrostatic simulation for 50,000 steps (20 seconds) shows that initializing with the BIPI algorithm significantly reduces noise compared to simulations without packing, as shown in Figure 13. The KE profiles in Figure 13 show disturbances near the wedge for simulations without particle initialization. For the BIPI algorithm, we used a 1% change in $TPD_{avg}$ as the stopping criterion for Step 2a and a 1% change in $|\nabla C|_{avg}$ for Step 2c. This criterion will be used throughout the article unless specific iteration numbers for Step 2a and Step 2c are mentioned.

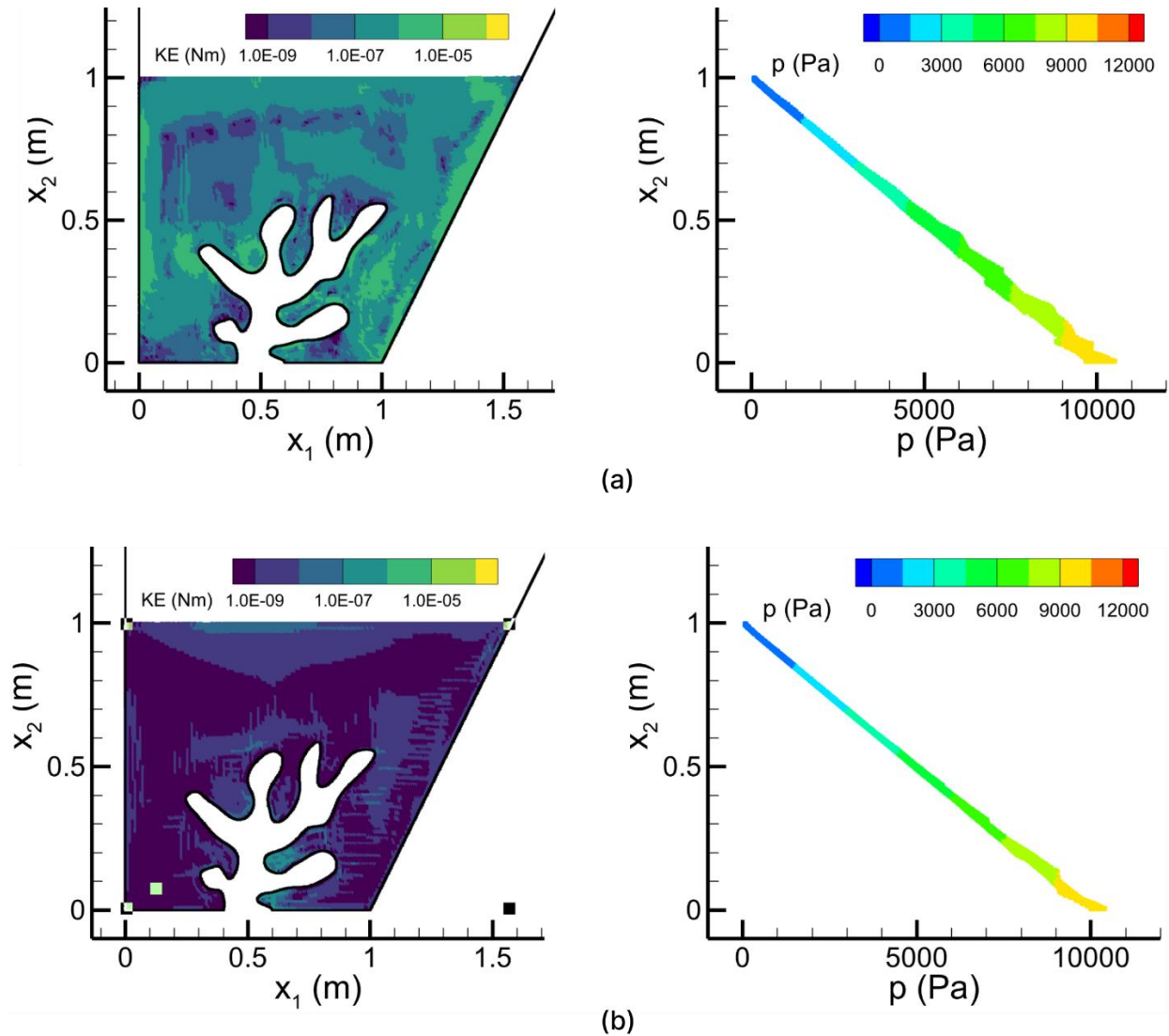

*Figure 15 Hydrostatic simulation of coral in tank after 10s, using particle spacing $dx_r = 0.01m$ and smoothing length to particle spacing ration $h/dx_r = 2$, for (a) Without BIPI pre-processing, (b) With BIPI preprocessing. Here KE is the Kinetic Energy, p is the pressure, $x_1$ is the horizontal coordinate and $x_2$ is the vertical coordinate.*

We also conduct convergence studies for different particle spacings, both with and without the BIPI algorithm as the initial configuration for the hydrostatic simulation. All simulations were run for 20 seconds. For all particle spacings, simulations that did not utilize the BIPI algorithm exhibit higher velocity field errors, as shown by the KE plot in Figure 14a. In contrast, simulations initialized with the BIPI algorithm demonstrate improved convergence rates, evidenced by lower KE values across different particle spacings. The use of the BIPI algorithm in hydrostatic simulations leads to a significant reduction in KE, highlighting its effectiveness.

To further examine how errors depend on the SPH formulations used, we conduct a comparative study by employing the Particle Consistent Gradient (PCG) formulation from [3–5] to evaluate the Laplacian in Equation (17) for the viscous term. We observe that the error behavior changes considerably when a different SPH operator is used (Figure14). Although the PCG formulation improves simulation accuracy near complex geometries, it is theoretically best suited when the fluid domain boundary is fully defined for evaluating the boundary integral term. While the presence of a free surface does not pose a problem, the interface between the free surface and wall requires special consideration. For this hydrostatic problem, we observe reduced noise in the KE graph when using the PCG term to evaluate the Laplacian. However, we do not recommend using the PCG term directly in dynamic free surface simulations where the interface between the free surface and wall is not explicitly modeled.

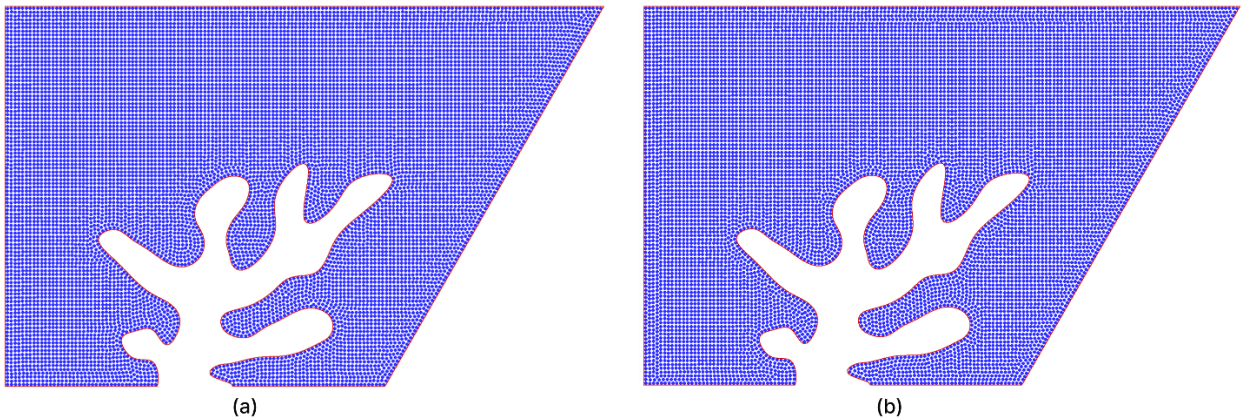

*Figure 16 Coral in a tank, packed with (a)1000 iterations of Step2a and (b) Step2a with 1% change in $TPD_{avg}$. In both cases Step2c is run until 1% change of $|\nabla C|_{avg}$*

The key takeaway is that while the BIPI algorithm improves simulation accuracy, the nature and behavior of long-term simulation errors are also dependent on the SPH formulation and explicit scheme used. Consequently, the BIPI algorithm should be viewed as a method to improve accuracy around complex boundaries, but it may not entirely eliminate simulation errors if the underlying SPH formulations used to model the physics give rise to errors.

Now, we consider a more complex coral geometry in a trapezoidal tank. The coral itself has various non-linear curves modeled with linear boundary elements. Modeling such complex geometry using ghost particles is challenging, but with BISPH, we can easily analyze these structures. As in the previous hydrostatic simulation, we set $F_{ext_2} = -9.81\ m/s^2$, $\rho_0 = 1000\ Kg/m^3$ and $\mu = 10\ Pa\ s$. The hydrostatic simulation is run for $20s$, both with and without the BIPI algorithm, for the initial particle configuration at $t = 0s$ and particle spacing $dx_r = 0.01m$. Again, we observe that the error in kinetic energy (KE) and the error in linear pressure variation are higher in simulations without the BIPI algorithm, as shown in Figure 15. The kinetic energy of particles, and thus the error in the velocity field, is significantly higher in simulations without the BIPI algorithm.

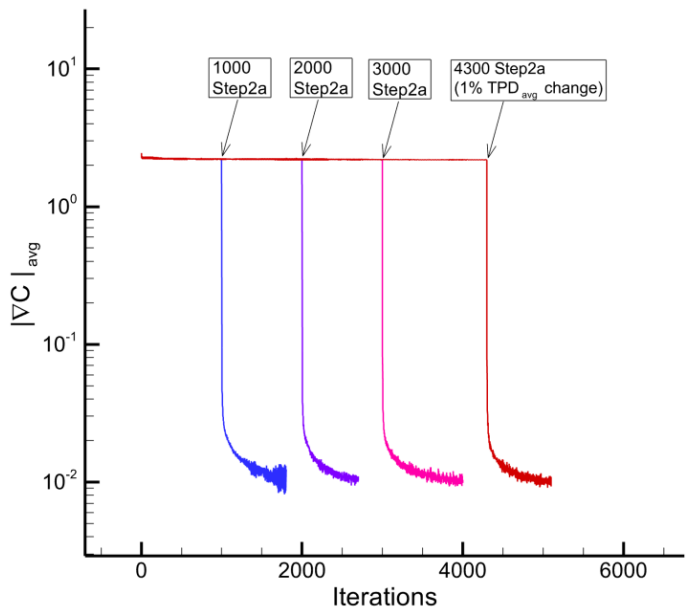

*Figure 17 Different stopping iterations of Step2a for Coral in Tank with particle spacing $dx_r = 0.01m$, showing average concentration gradient plot.*

For particle spacing $dx_r = 0.01m$ in the coral-in-tank case, applying the BIPI algorithm with a criterion of 1% change in $TPD_{avg}$ for Step 2a and 1% change in $|\nabla C|_{avg}$ for Step 2c results in 4300 iterations for Step 2a and 800 iterations for Step 2c. Notably, the layer of particles adjacent to the boundary conforms to the boundary shape much earlier. To investigate the effect of reducing the number of iterations in Step 2a while maintaining approximately the same number of iterations in Step 2c to achieve the 1% change in $|\nabla C|_{avg}$, we conduct additional tests. Figure 16 shows that with only 1000 iterations of Step 2a and 800 iterations of Step 2c, the particle configuration aligns well with the boundary, and the final $|\nabla C|_{avg}$ at the end of Step 2c is consistent, as observed in blue line of Figure 17.

Further, we perform hydrostatic simulations for the coral-in-tank scenario for different iterations of Step 2a in the BIPI algorithm over 20s, using particle spacings of 0.01m and 0.005m. We analyze the evolution of KE for these simulations in Figure 18, comparing them with hydrostatic simulations without the BIPI algorithm for both particle spacings. We make three observations from these results: first, in all cases, simulations without BIPI have higher KE and thus more error in the velocity field compared to those with the BIPI algorithm; second, varying the number of iterations in Step 2a does not significantly affect accuracy or stability, indicating that Step 2a can be stopped earlier than a 1% change in $TPD_{avg}$ if particles sufficiently follow the boundary. Third, for smaller particle spacing, even with the BIPI algorithm, some disturbances in the velocity field are observed due to the curvature radius of the coral structure approaching particle spacing; however, these errors decrease, and stability improves by further decreasing particle spacing. It is important to note that the curves in Figure 18 are on a log scale, and the magnitude of errors observed remain much lower than those seen in simulations without the BIPI algorithm.

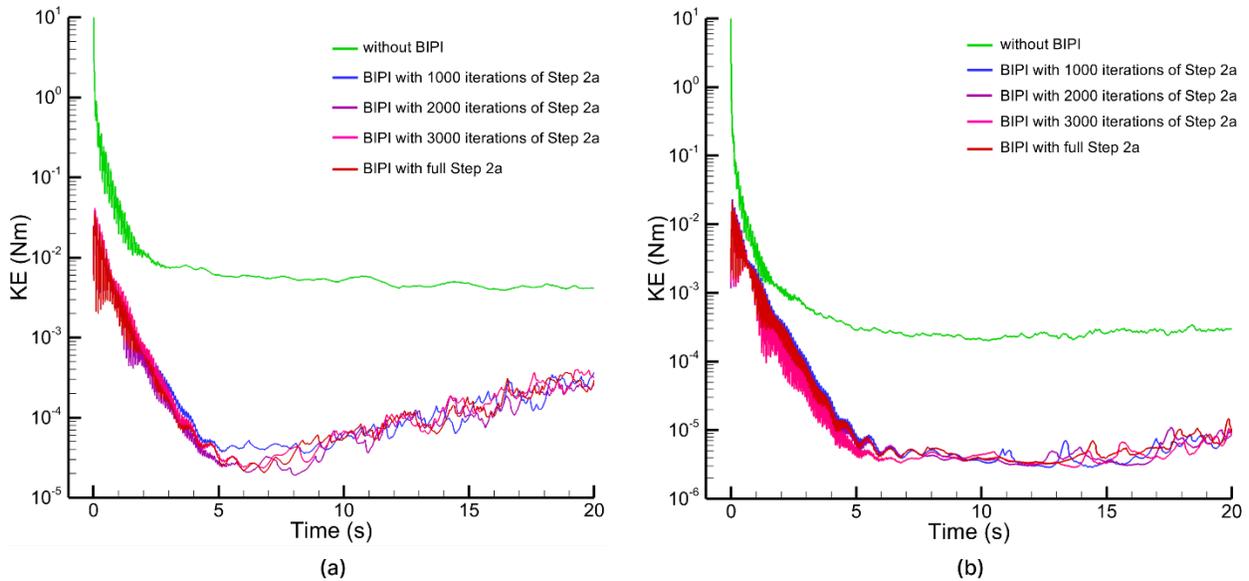

*Figure 18 Comparison of Kinetic Energy (KE) evolution hydrostatic simulation of coral structure with different number of BIPI step2a iterations and without BIPI algorithm. The evolution of KE is compared for two different particle spacing (a) $dx_r = 0.01m$ and (b) $dx_r = 0.005m$*

*4.2.2 Dynamic Flow Simulation*

To demonstrate the effect of the packing algorithm on dynamic flow simulations, we consider two cases: a flow over periodic array of cylinders between parallel plates and the evolution of an elliptical drop. In both cases, the simulation begins with a non-zero initial condition to show the impact of the BIPI algorithm. Starting with a non-zero initial condition draw attention to larger errors in SPH approximations like (8), which are $C^0$ particle consistent for untruncated kernels and thus can only accurately approximate constant fields (and its derivatives). Additionally, many practical applications commence with non-zero initial boundaries. Designing a ship hull, as demonstrated by Colagrossi et al. [9], reveals the advantage of a particle packing algorithm in such cases. Since our current code is tested only for fluid applications without structure interaction, we demonstrate BIPI algorithm with a different flow example case.

First, we simulate fluid flow around a cylinder array like that used by [17] but between two parallel plates. The fluid is accelerated by a force of $F_{ext_1} = 100 \ m/s^2$, horizontally, between the two parallel plates. At steady state, this yields a Reynolds number of $Re = \frac{\rho_0 \bar{v} r_{cyl}}{\mu} \approx 9.8$, calculated using the average flow velocity ($\bar{v}$) within one smoothing kernel diameter adjacent to the periodic boundary. Here, $\rho_0 = 1000 \ kg/m^3$, $r_{cyl} = 0.02m$, and $\mu = 10 \ Pa \ s$. We consider three SPH simulation scenarios: (i) without BIPI preprocessing, (ii) with BIPI preprocessing for 100 iterations of both Step 2a and Step 2c, and (iii) with BIPI preprocessing for 400 iterations of Step 2a and 300 iterations of Step 2c.

Rather than starting the simulation from rest, we first obtain steady-state velocity and pressure data from a commercial incompressible solver (STAR-CCM+). This steady-state data serves as the initial condition for the SPH simulation. To model the pressure field in WCSPH, we use the minimum pressure from the steady-state results as background pressure ($P_B$) in the equation of state (18), which helps maintain positive pressure throughout the simulation and prevents tensile instability. To improve the accuracy of the Laplacian formulation used to calculate viscous stress, we use the Particle Consistent Gradient (PCG) formulation to evaluate $\nabla \cdot \nabla v_a$ in (17). Ideally, velocity and pressure should remain unchanged once the SPH simulation begins, as steady-state conditions have already been achieved.

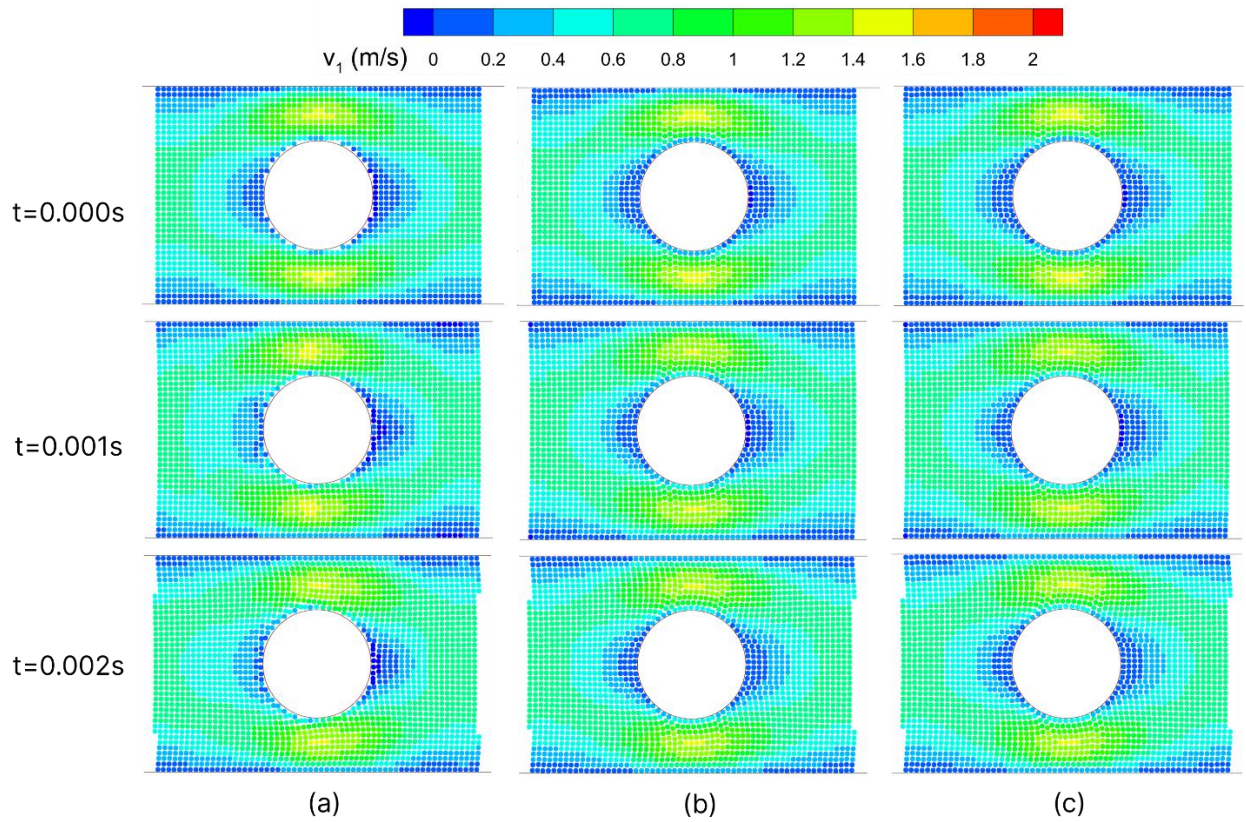

*Figure 19 Flow over cylinder array at timestamps 0s, 0.001s, and 0.002s, showing horizontal velocity field ($v_1$): (a) without BIPI, (b) with BIPI preprocessing for 100 iterations of Step 2a and Step 2c, (c) with BIPI preprocessing for 400 iterations of Step 2a and 300 iterations of Step 2c.*

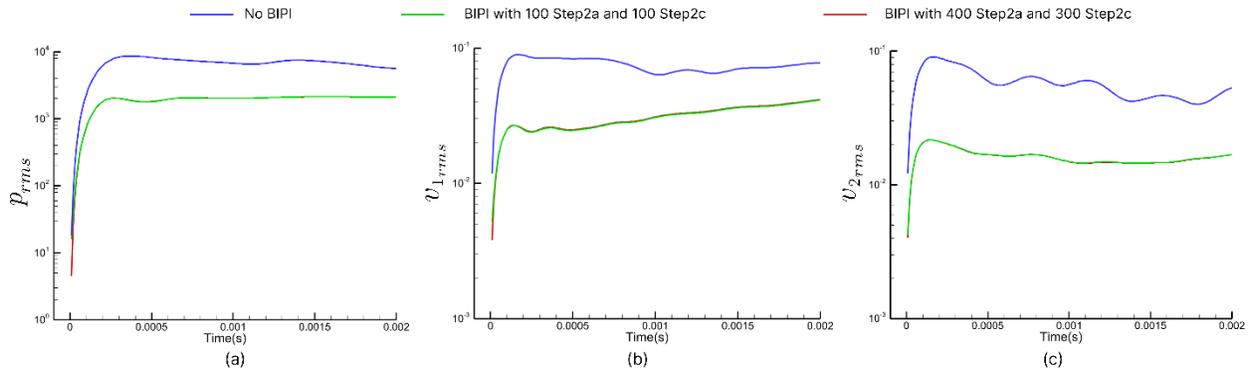

*Figure 20 Evolution of initial Root Mean Square(RMS) errors for with and without BIPI initialized simulation of flow over cylinder array, with plots of (a) RMS error in pressure($p_{rms}$), (b)RMS error in horizontal velocity ($v_{1_{rms}}$), and (c)RMS error in vertical velocity ($v_{2_{rms}}$)*

Due to the weakly compressible nature of the SPH equations and the initial conditions derived from the incompressible solver's steady state, in addition to discretization errors of SPH formulation, some deviation in SPH simulation are inevitable. These errors could be reduced further by adjusting the density diffusion coefficient, adding artificial viscosity, using PST, or employing Arbitrary Lagrangian-Eulerian (ALE) equations. However, our focus is not to use the best WCSPH model but rather to illustrate the impact of the BIPI algorithm on an SPH simulation. Upon comparing simulations with and without the BIPI algorithm, in Figure 19, we observe higher initial errors in simulations without BIPI, which propagate until a steady state, corresponding to the WCSPH simulation's steady state, is eventually reached. Figure 20 shows how these errors propagate with Root Mean Square (RMS) errors for pressure, horizontal velocity, and vertical velocity labeled as $p_{rms}$, $v_{1_{rms}}$ and $v_{2_{rms}}$, respectively. The RMS errors are calculated for all N particles in the simulation, using the values of initial pressure ($p^{(0)}$), initial horizontal velocity ($v_1^{(0)}$) and initial vertical velocity ($v_2^{(0)}$) as the reference parameter for RMS error calculation.

Furthermore, we observe that the pressure field disturbance in simulations without the BIPI algorithm remains higher and persists longer, as shown in Figures 20a and 21a. Although some of these errors stem from acoustic noise due to weak compressibility, simulations initialized with the BIPI scheme show smaller and comparable errors to the initial conditions. Notably, across Figures 19-21, only 100 iterations of Step 2c yield accuracies comparable to those achieved after 300 iterations, highlighting the importance of boundary shape conformity achieved by Step 2a. Only a few iterations of Step 2c are required to further redistribute internal particles, making additional reduction of the concentration gradient unnecessary to improve simulation results.

Next, we consider the case of elliptical drop evolution, by considering a circle of radius R, subjected to an initial velocity field ($\boldsymbol{v}^{(0)}$) and pressure field ($p^{(0)}$), given by,

$$\begin{aligned} v_1^{(0)} &= A_0\, x_1 \\ v_2^{(0)} &= -A_0\, x_2 \\ p^{(0)} &= \frac{\rho_0 A_0^2}{2}\left(R_0^2 - (x_1^2 + x_2^2)\right) \end{aligned} \tag{22}$$

Where $x_1$ is the horizontal coordinate, $x_2$ is the vertical coordinate, $R_0$ is the initial radius of the circle, and $A_0 t$ gives the non-dimensional time parameter (where $t$ is time). For an incompressible, inviscid flow, and analytical solution can be derived which allows us to predict the evolution of initial 2D circular drop to 2D elliptical shape [9,22] . We compare the evolution of the drop at $A_0 t = 2$, for a configuration simulated without BIPI and other simulated after packing with BIPI and notice that the simulation after BIPI follows the analytical drop boundary better than the simulation without BIPI, as seen in Figure 22.

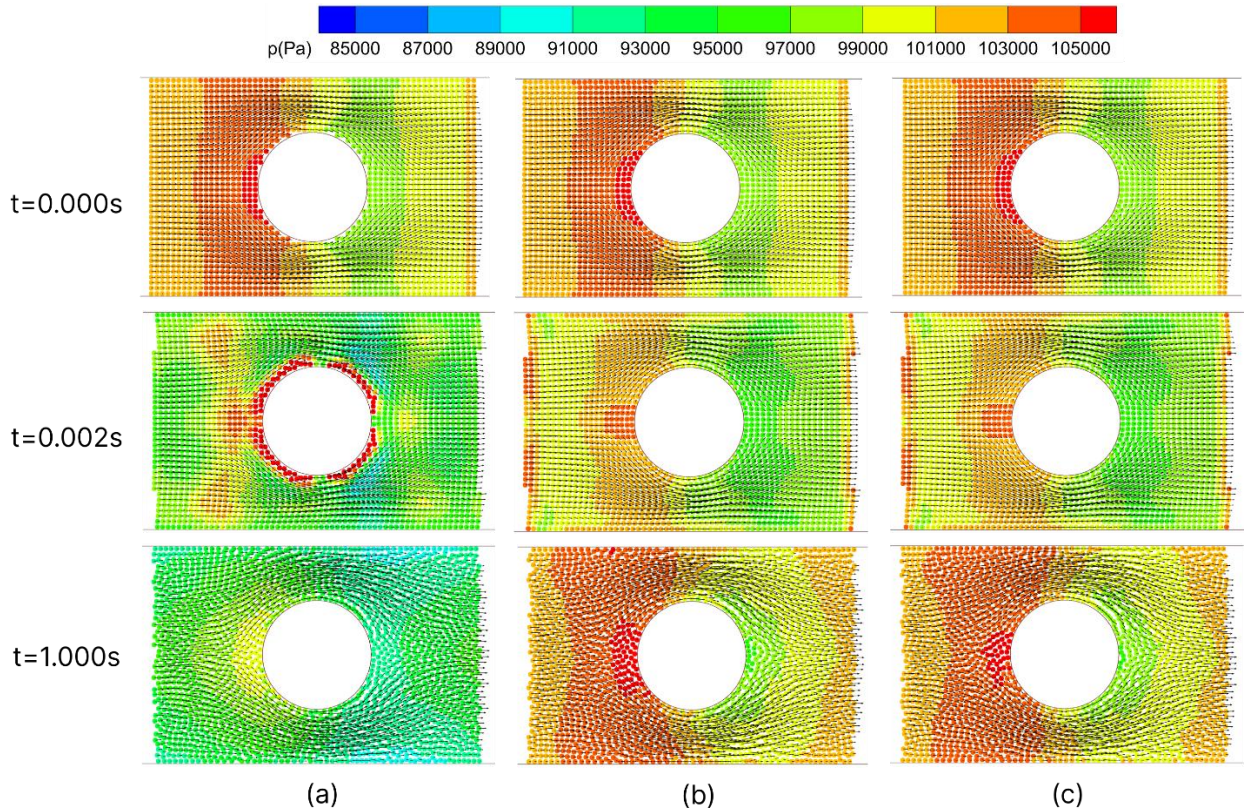

*Figure 21 Flow over cylinder array at timestamps 0s, 0.002s, and 1s, showing pressure field (p) and velocity vectors($\boldsymbol{v}$): (a) without BIPI, (b) with BIPI preprocessing for 100 iterations of Step 2a and Step 2c, (c) with BIPI preprocessing for 400 iterations of Step 2a and 300 iterations of Step 2c.*

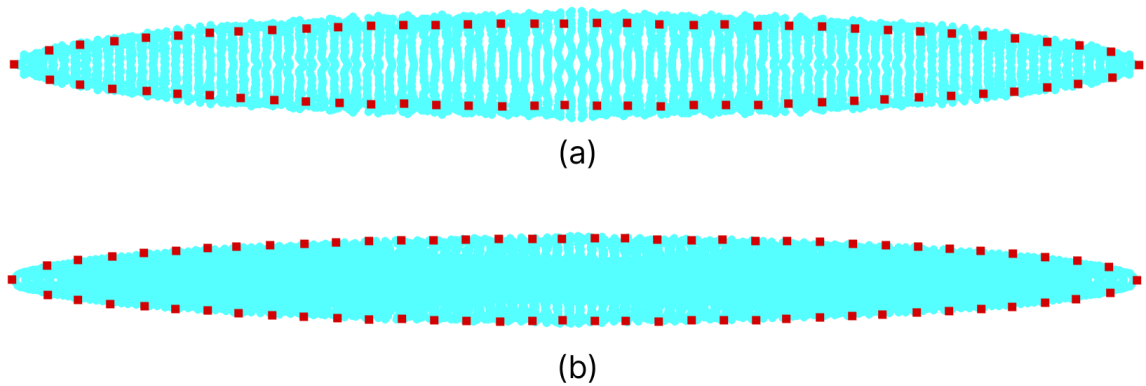

*Figure 22 Elliptic drop evolution (a) without BIPI and (b) with BIPI*

## 4.3 Time complexity of BIPI algorithm

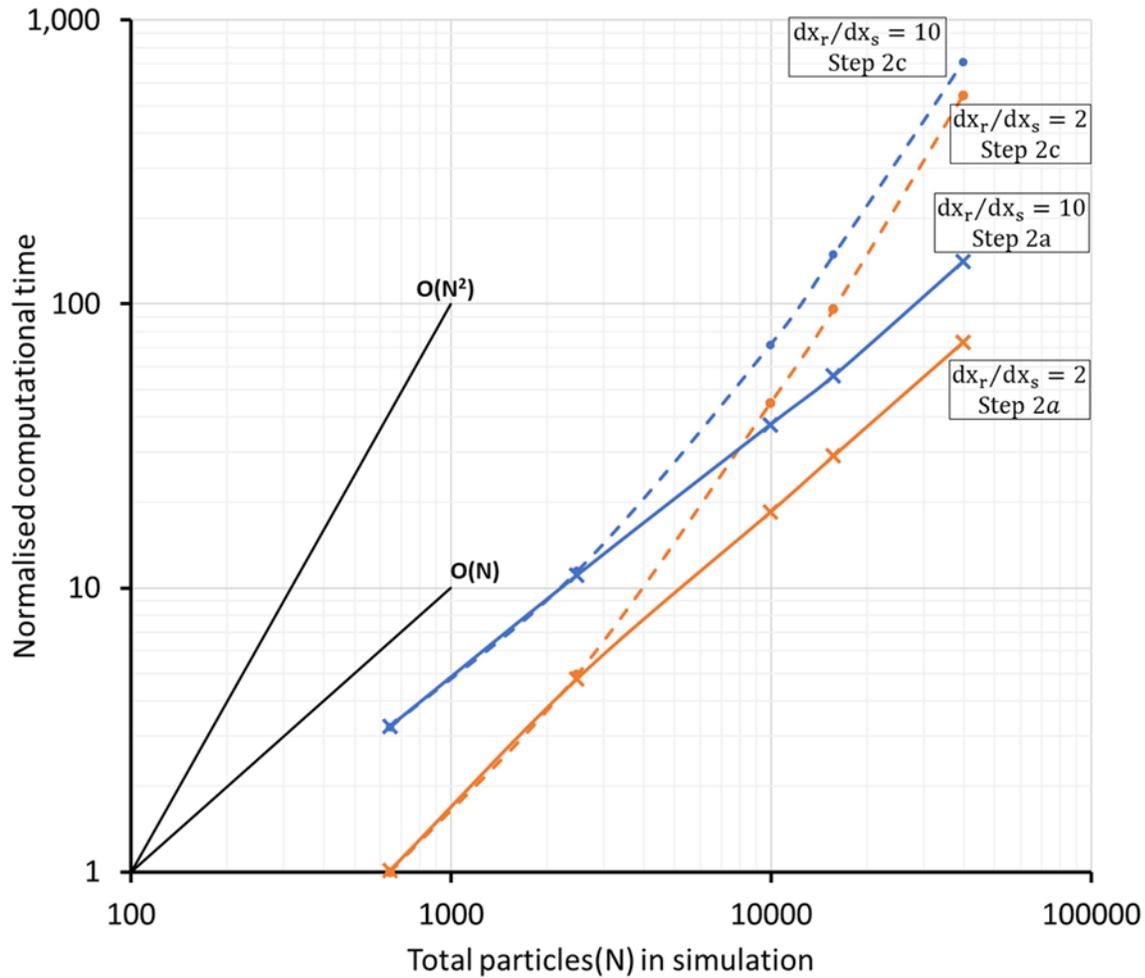

*Figure 23 Computational time complexity of each iteration in Step2a and Step2c, for two different particle spacing to boundary element length ratios ($dx_r/dx_s$).*

In Sections 4.1 and 4.2, we observe that Step 2a plays a crucial role in distributing particles near the boundary to follow the boundary shape, while Step 2c reduces the concentration gradient by further redistributing particles away from the boundary. Once particles near the boundary align with its shape (using Step 2a), only a few iterations of Step 2c are typically needed to further reduce the concentration gradient. Although running Step 2c for an extended period continues to lower the concentration gradient, finding a global minimum is neither guaranteed nor necessary. The purpose of the BIPI algorithm is simply to reduce errors arising from improper particle distribution near the boundary at the onset of the simulation.

A key component of the BIPI algorithm is Step 2a. Not only does it enable effective alignment of particles with the boundary shape before redistributing interior particles, but it also reduces computation time by focusing only on a small subset of particles near the domain boundary. To qualitatively demonstrate this, we analyze the time complexity of Step 2a and Step 2c per iteration for the trapezoidal geometry case of Section 4.1. For this comparison, we employ a direct neighbor search algorithm, which represents a worst-case search algorithm to determine particle pairs. Even under these conditions, as the number of particles is increased to model the same geometry, Step 2a exhibits nearly linear behavior, while Step 2c follows approximately quadratic behavior (see Figure 23). This result is expected, as Step 2a operates exclusively on particles adjacent to the boundary. Additional performance gains can be expected by employing more efficient neighbor search techniques, such as linked-list or tree-search algorithms[34].

Moreover, the number of iterations in Step 2c is considerably lower than in Step 2a, resulting in significant computational time savings for particle packing. Without the BIPI algorithm, simulations would require solving the physics equations with a high damping term, as suggested by Monaghan [9,22], to achieve a suitable initial particle distribution. This alternative approach tends to be slower, as it involves solving equations with a time complexity on par with Step 2c, and often complicated by higher-order terms and time step-dependent simulation constraints. By implementing the BIPI algorithm, particularly through the redistribution of particles near the boundary in Step 2a, we can achieve substantial reductions in overall computational time.

## 5  Future Work and Current Limitations

While we present the only particle packing algorithm that can be used for boundary integral SPH models without relying on virtual particle layers, much work remains to be done to test and extend this model to other SPH scenarios. Although we hope to counter the following current limitations in future work, we hope these issues will allow other researchers to further extend the work into these and other assumptions we may have overlooked.

The first issue lies with our current tests, which have been limited to the Wendland kernel. This is due to the use of semi-analytical calculations of $\gamma_a$ and $\nabla\gamma_{as}$, which are commonly employed in existing boundary integral research. While it is possible to calculate the boundary term using numerical integration and apply it to different kernels, we have not yet adopted this approach. This is because both numerical integration and the application of the boundary integral formulation across a wide range of smoothing kernels remain largely unexplored in boundary integral research. In theory, all smoothing kernels used in traditional SPH research should be extendable to BISPH for calculating $\gamma_a$ and $\nabla\gamma_{as}$ and by extension, applicable to the BIPI algorithm. However, current literature has only utilized a limited selection of kernels in Boundary Integral SPH, and more extensive research is needed to explore the use of various kernels in 2D and 3D, including those with higher-order convergence[35].

The second issue concerns the extensive testing conducted primarily on 2D models in this study. Although the BIPI formulations for both 2D and 3D cases are derived and proposed in Section 3, the algorithm is only demonstrated for 2D scenarios. Some extensions, such as splitting a line element into additional line segments to ensure $1 < dx_r/s$, are relatively straightforward when applied to 3D cases by splitting large triangle elements into multiple triangles using algorithms like Delaunay triangulation. However, other aspects, such as implementing semi-analytical values for $\gamma_a$ and $\nabla\gamma_{as}$ in 3D, require additional work, as highlighted in the studies by Chiron et al [21]. We also anticipate that Step2a in 3D will further improve computational speed, benefiting users who apply our formulations to 3D cases.

The third issue concerns the assumption of uniform particle resolution during the derivation. For simulations involving multi-resolution particle distribution, updates would be required to the boundary force in (14), the upper bound in (15), and the value of $D_a$. While we believe that much of our algorithm can be adapted to account for varying particle radii and smoothing length - differing from the current assumption of a uniform smoothing length and uniform particle radius - extensive testing is necessary. We have not yet conducted tests for simulations with multi-particle resolution, and such cases are of practical importance that could greatly benefit from particle packing algorithms, hence requiring further investigation.

Finally, we make an additional note regarding the use of damping terms in particle packing algorithms. Most particle packing algorithms in the literature utilize a damping term for velocity, which helps particles settle by reducing the system's energy. The challenge we encountered when using this damping term primarily involved particles near the boundary whose kernels are truncated. Unlike traditional methods, our algorithm does not require generating virtual particle layers to complete kernel support near the boundary. In theory, however, once Step 2a is applied and kernel-truncated particles near the boundary are frozen in Step 2b, and using a slightly higher values of $k_a$ and $k_b$, the damping term could be employed in Step 2c. This approach would allow for the effective use of frozen real particles near the boundary as pseudo virtual particle layer for packing, without the need to create external ghost particles during initialization—a process that can be challenging for complex shapes. However, it is uncertain whether this method would be computationally more efficient than the algorithm we propose here, and further investigation is needed to determine its effectiveness.

# 6 Conclusion

The Boundary Integral based Particle Initialization (BIPI) algorithm proposed here represents a significant advancement in particle initialization techniques for Smooth Particle Hydrodynamics (SPH) simulations in the following ways:

Firstly, BIPI revolutionizes particle initialization by eliminating the need for virtual particle layers, directly utilizing boundary information through boundary integrals. This approach streamlines the modeling process for complex boundaries.

Secondly, unlike traditional initialization models that directly solve the momentum equation, BIPI simplifies the process through a particle shifting like algorithm. This algorithm is applied to both interior particles and those near the boundary, with the latter utilizing boundary integral formulation to accommodate truncated kernels.

Moreover, BIPI ensures a uniform particle configuration by reducing concentration gradient errors. This optimization improves simulation accuracy and stability, which is crucial for achieving reliable results. The fluid flow simulation studies in this work also demonstrate that even a few iterations of BIPI are sufficient to noticeably improve the simulation outcomes.

Additionally, by implementing a boundary force, BIPI addresses the challenge of artificial compression of particle volume near the boundary. This prevents undesirable distortions, maintains the particle distribution's integrity, and allows faster convergence to conditionally stable particle configuration.

Lastly, BIPI optimizes computational efficiency by prioritizing particles fitting along the boundary before redistributing particles throughout the domain. This strategic approach minimizes the computational overhead of the particle initialization process.